\documentclass[fleqn,usenatbib]{mnras}

\usepackage[T1]{fontenc}

\DeclareRobustCommand{\VAN}[3]{#2}
\let\VANthebibliography\thebibliography
\def\thebibliography{\DeclareRobustCommand{\VAN}[3]{##3}\VANthebibliography}
\providecommand{\abs}[1]{\lvert#1\rvert}

\usepackage[dvipsnames]{xcolor}

\usepackage{graphicx}	
\usepackage{amsmath}	
\usepackage{bm}
\usepackage{caption}
\usepackage{subcaption}

\usepackage[dvipsnames]{xcolor}

\usepackage{newtxtext,newtxmath}
\title[Resonant drag instabilities in 3D discs]{Planetary waves can activate resonant drag instabilities in 3D dusty gaseous discs}

\author[R. O. Chametla et al.]{
Raúl O. Chametla$^{1},$\thanks{E-mail: raul@sirrah.troja.mff.cuni.cz (ROC)}
Yasuhiro Hasegawa$^{2}$
and Gennaro D'Angelo$^{3}$
\\
$^{1}$Charles University, Faculty of Mathematics and Physics, Astronomical Institute, V Hole\v{s}ovi\v{c}k\'ach 747/2, 180 00 Prague 8, Czech Republic\\
$^{2}$Jet Propulsion Laboratory, California Institute of Technology, Pasadena, CA 91109, USA\\
$^{3}$Theoretical Division, Los Alamos National Laboratory, Los Alamos, NM 87545, USA\\
}

\date{Accepted XXX. Received YYY; in original form ZZZ}

\pubyear{2023}

\begin{document}
\label{firstpage}
\pagerange{\pageref{firstpage}--\pageref{lastpage}}
\maketitle

\begin{abstract}
Resonant Drag Instabilities (RDIs) in protoplanetary discs are driven by the aerodynamic back-reaction of dust on gas and occur when the relative dust-gas motion 
resonate with a wave mode intrinsic to the gas fluid. Axisymmetric models indicate that the RDI generates filamentary perturbations, leading to grain clumping and 
planetesimal formation. Motivated by these findings, we investigate the dust-gas interaction in a non-axisymmetric inviscid protoplanetary disc with an embedded 
low-mass planet ($M_{\mathrm{p}}\in[0.3, 3] M_\oplus$, here $M_\oplus$ is the Earth mass). We conduct global 3D high-resolution two-fluid simulations, with 
the dust being parametrized by the Stokes number $\mathrm{St}\in[0.01,0.5]$. We find that planetary waves (PWs; also known as Rossby waves), which propagate 
along the downstream separatrices of the horseshoe region, resonate with the streaming motion and trigger the RDI.
The consequent development of a global-scale filamentary dust distribution does not sensitively depend on the Stokes number, 
nor does it depend on the fast dust settling that takes place in an inviscid disc. The rapid onset of this instability, which is comparable to the dynamical orbital time-scale, suppresses the formation of asymmetric structures
in the dust in the vicinity of the planet (such as dust voids and filaments).
Additionally, we find that the dust feedback enables buoyancy resonances in an otherwise non-buoyant (globally isothermal) disc. Therefore, our results provide the first numerical evidence of RDIs generation driven by planetary waves.
\end{abstract}




\section{Introduction}
Recent studies of dust grains streaming through a gaseous medium with a constant relative velocity $\mathbf{w}_\mathrm{s}\equiv\mathbf{u}-\mathbf{v}$ (where $\mathbf{u}$ and $\mathbf{v}$ are the dust and gas velocities, respectively) have shown
that the coupled fluids can become unstable by means of the resonant drag instability \citep[RDI;][]{Squire_Hopkins2018,SH2018,SH2020Physicalmodels}. The main condition that must be met for the RDI to occur is
$\mathbf{k}\cdot\mathbf{w_\mathrm{s}}=\omega_\mathrm{gas}(\mathbf{k})$ \citep{Squire_Hopkins2018}, where $\omega_\mathrm{gas}(\mathbf{k})$ is the frequency of an intrinsic wave mode of the gas (existing without any dust)
and $\mathbf{k}$ is the mode's wavenumber. In other words, the RDI deveolps whenever any component of
$\mathbf{w}_\mathrm{s}$ matches the phase velocity of a wave traveling through the gas.

A suitable environment for the RDIs to develop is in protoplanetary discs \citep[][]{Squire_Hopkins2018,SH2020Physicalmodels}, which are composed of gas and a small mass fraction ($\sim1$$\%$) of dust. Indeed, a variety of RDI-driven dust-gas instabilities have been found to influence the structure of protoplanetary discs, for instance: the Streaming Instability \citep[SI;][]{YG2005,JBL2011}, the Disc Settling Instability \citep[DSI;][]{SH2018,Lin2019,Krapp_etal2020}, the Brunt-Väisälä RDI (BV-RDI), and the Magnetosonic RDI \citep[M-RDI; see][and references therein]{Squire_Hopkins2018,SH2020Physicalmodels}. Furthermore, all these instabilities play an important role in forming planetesimals because they can accumulate
dust particles and locally increase their concentrations, thus allowing for their clumping by gravitational collapse.
While these instabilities can easily occur in some disc regions, other regions might require the dust-to-gas mass ratio
to be close to unity \citep[$\rho_\mathrm{d}/\rho_\mathrm{g}\approx 1$;][]{YG2005,JBL2011} to eventually form planetesimals.
A substantial increase of the dust-to-gas mass ratio can be achieved, for instance, in the disc midplane when the dust vertically settles.
However, the settling time-scale depends on the grain size as well as on the aerodynamic dust feedback and the
overall level of gas turbulence, which can speed up settling, or slow it down.

Here, we focus on the possibility of activating the RDI in the presence of planetary waves (PWs) generated by a disc-embedded planet. These waves, more commonly known as Rossby waves in protoplanetary discs, are excited near density or vortensity maxima \citep{Lovelaceetal1999,Lietal2000,Lietal2001}.
Such vortensity perturbations commonly exist in the horseshoe region of a planet and within its Hill sphere, where gas
accumulates in the potential well of the planet. Our working hypothesis is that, when the PWs are present,
they can resonate with the streaming motion of the gas and dust, thus making the mixture unstable triggering SI-like\footnote{SI itself belongs to the set of RDIs \citep[see][]{SH2020Physicalmodels}.} filamentary structures.
Since the existing framework of the RDI theory is not yet advanced enough to confirm our hypothesis analytically, we do so by performing 3D high-resolution two-fluid simulations of low-mass planets embedded in the protoplanetary disc. We find SI-like disturbances in the dust density confirming our assumption since the filamentary disturbances of the dust only occur when the aerodynamic feedback of the dust on the gas is taken into account,
as required by the RDI mechanism.

The paper is organized as follows. In Section~\ref{sec:initial} we present the numerical model used in our 3D dust-gas hydrodynamic simulations.
In Section~\ref{sec:results}, the results of our numerical simulations are described. The possible implications of the instability are discussed
in Section~\ref{sec:discussion}. Concluding remarks can be found in Section~\ref{sec:conclusions}.

\begin{table}
\caption{List of numerical simulations and their parameters.}
    \label{tab:simulations}

\begin{tabular}{ p{1.9cm} p{1.1cm} p{1.9cm} p{0.8cm}}

 \hline
 Run & $M_\mathrm{p}[M_\oplus]$ & $\mathrm{St}$ & Feedback \\
 \hline
 RDISt0Mp03 & 0.3 & 0.0  & No\\
 RDISt0Mp1 & 1.0 & 0.0  & No\\
 RDISt0Mp3 & 3.0 & 0.0  & No\\
 RDISt0$\_$05Mp1 & 1.0 & 0.05   & Yes\\
 RDISt0$\_$2Mp0 & 0.0 & 0.2  & Yes\\
 RDISt0$\_$2Mp1 & 1.0 & 0.2   & Yes\\
 RDISt0$\_$5Mp1 & 1.0 & 0.5   & Yes\\
 RDISt0$\_$2Mp03 & 0.3 & 0.2   & Yes\\
 RDISt0$\_$2Mp03a & 0.3 & 0.2   & No\\
 RDISt0$\_$2Mp3 & 3.0 & 0.2   & Yes\\
 RDISt0$\_$01$^{H_d}$Mp1 & 1.0 & 0.01   & Yes\\ 
 \hline
\end{tabular}
\end{table}

\section{Dusty disc-planet model}
\label{sec:initial}
Here, we describe different components of our physical model: the gas disc, the dust disc, the gravitational potential.
We also provide some information on the code used to solve the set of hydrodynamics equations.
\subsection{Governing equations}
\label{sec:gas}
The problem is studied in spherical coordinates $(r,\theta,\phi)$, where $r$ is the radial distance from the star, $\theta$ is the colatitude ($\theta=\pi/2$ at the midplane of the disc), and $\phi$ is the azimuthal angle.
We consider a 3D, non-self-gravitating, inviscid gas-dust disc whose evolution 
is governed by the following equations:

\begin{equation}
\partial_t\rho_\mathrm{g}+\nabla\cdot(\rho_\mathrm{g} \mathbf{v})=0,
 \label{eq:gas_cont}
\end{equation}
\begin{equation}
\partial_t\rho_\mathrm{d}+\nabla\cdot(\rho_\mathrm{d} \mathbf{u})=0,
 \label{eq:dust_cont}
\end{equation}
\begin{equation}
\partial_t(\rho_\mathrm{g}\mathbf{v})+\nabla\cdot(\rho_\mathrm{g}\mathbf{v}\otimes\mathbf{v})=-\nabla p -\rho_\mathrm{g}\nabla\Phi-\mathbf{f}_\mathrm{d},
 \label{eq:gas_mom}
\end{equation}
\begin{equation}
\partial_t(\rho_\mathrm{d}\mathbf{u})+\nabla\cdot(\rho_\mathrm{d}\mathbf{u}\otimes\mathbf{u})=-\rho_\mathrm{d}\nabla\Phi+\mathbf{f}_\mathrm{d},
 \label{eq:dust_mom}
\end{equation}
where $\rho_\mathrm{g}$, $\rho_\mathrm{d}$, $\mathbf{v}$, and $\mathbf{u}$ denote the gas and dust densities, and the gas and dust velocities, respectively. Furthermore, $\Phi$ denotes the gravitational
potential, $\mathsf{I}$ is the unit tensor, and $p$ is the gas pressure. For the latter, we consider a globally isothermal
equation of state
\begin{equation}
p=c_{\mathrm{s}}^2\rho_\mathrm{g},
 \label{eq:pressure}
\end{equation}
where $c_\mathrm{s}$ is the isothermal sound speed. The sound speed sets the gas pressure scale-height $H_\mathrm{g}=c_{\mathrm{s}}/\Omega_{\mathrm{K}}$, where $\Omega_{\mathrm{K}}=\sqrt{GM_\star/r^3}$ is the Keplerian angular velocity, $G$ is the gravitational constant,
and $M_{\star}$ is the mass of the central star (here we adopt $M_{\star}=1\,\mathrm{[c.u]}=1M_\odot$, with $M_\odot$ the solar mass).

In Equations~(\ref{eq:gas_mom}) and (\ref{eq:dust_mom}), the last term on the right-hand side is the drag force per unit mass given by
\begin{equation}
    \mathbf{f}_\mathrm{d}=\Omega_{\mathrm{K}}\mathrm{St}^{-1}(\mathbf{v}-\mathbf{u}) \,,
	\label{eq:dragforce}
\end{equation}
where the Stokes number, $\mathrm{St}$, is a dimensionless number that measures
the coupling timescale of dust and gas velocities. The Stokes number depends on
the drag coefficient of the dust particles and, in the Epstein regime 
\citep[of specular reflection, see][]{DP2015}, it can be approximated to
\begin{equation}
\mathrm{St}=\sqrt{\frac{\pi}{8}}\frac{r_\mathrm{m}\rho_\mathrm{m}\Omega}{\rho_\mathrm{g}c_s},
    \label{eq:St}
\end{equation}
where $r_{\mathrm{m}}$ is the grain radius, $\rho_{\mathrm{m}}$ is the dust material density
and $\Omega$ is the angular frequency, respectively.


\begin{figure}
  \centering
    \includegraphics[width=\columnwidth]{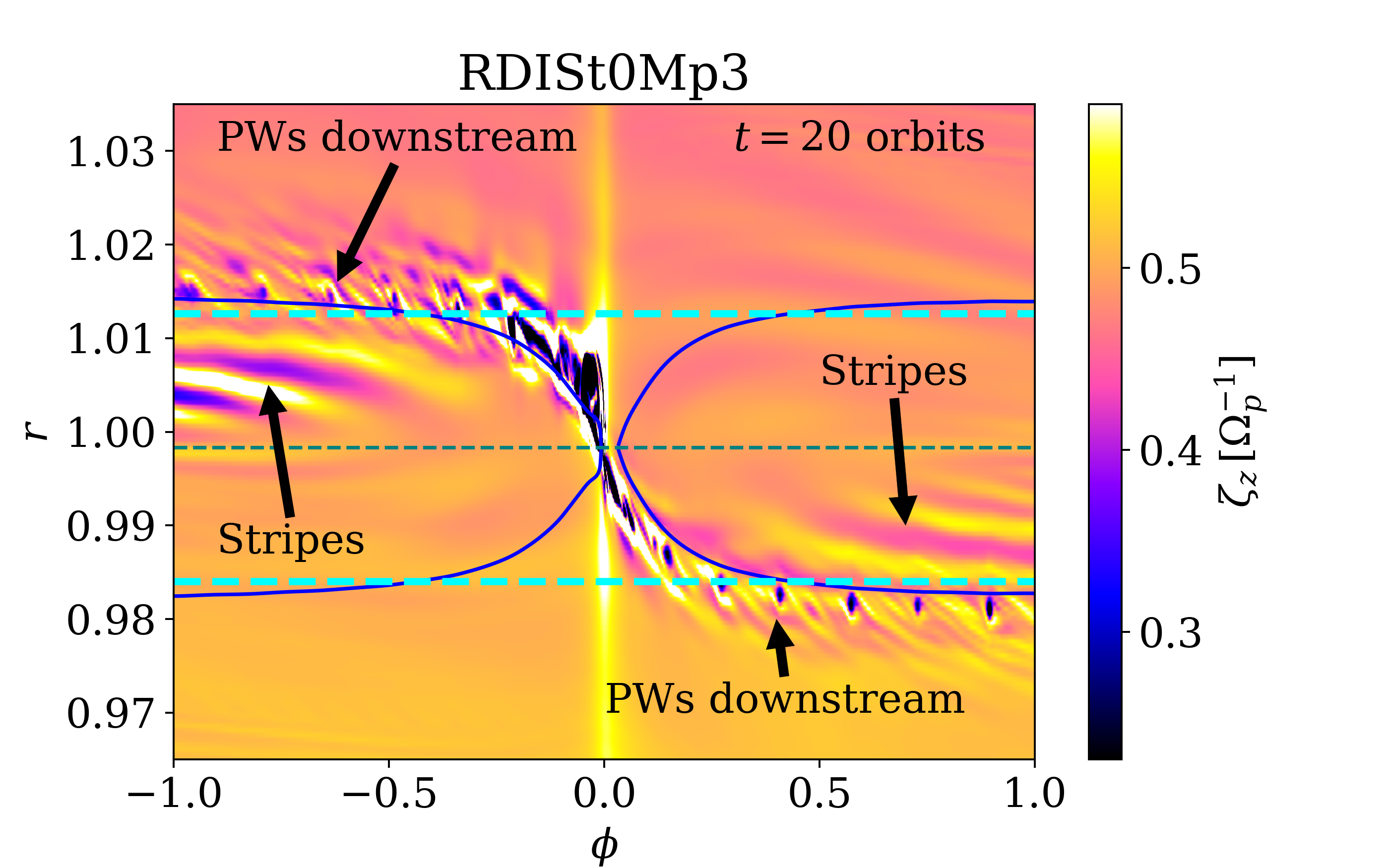}
  \caption{2D map of the vorticity in the midplane ($\theta=\pi/2$), resulting from model RDISt0Mp3, at $t=20$ orbits.
  We mark the locations
  of (i) the planetary waves propagating mainly along the downstream separatrices of the horseshoe region and (ii) the stripes confined
  within the horseshoe region. The stripes originate from the vertical oscillations of gas parcels upon executing their U-turns 
  (see Fig.~\ref{fig:stream3d}). The solid blue and dashed cyan lines represent, respectively, 
  the separatrices and the width of the horseshoe region predicted by Eq.~(\ref{eq:xs}).
  The dotted green line marks the corotation radius $r_\mathrm{c}$.}
  \label{fig:vortMS}
\end{figure}

\subsection{Setup}
\label{sec:dusty_model}

The surface density is chosen to be a power law
  \begin{equation}
    \Sigma_\mathrm{g}(r)=\Sigma_0\left(\frac{r}{r_\mathrm{p}}\right)^{-\alpha},
    \label{eq:Sigma}
  \end{equation}
  we set the slope\footnote{Note that, although this value of the slope generates an initial vortensity gradient in the disc, which could affect the vortensity in the horseshoe region, we verified our results against a slope $\alpha=1.5$ (not shown here) which means a null radial gradient of initial vortensity.} $\alpha=0.5$ and the surface density $\Sigma_0=6.05\times10^{-4}\,[\mathrm{c.u}]\simeq200\,\mathrm{g}/\mathrm{cm^2}$, at the orbital radius of the planet $r=r_\mathrm{p}=5.2$a.u. We initialize the gas velocity components in a way similar to \citet[][see their appendix A]{MB2016}, setting $v_r=v_\theta=0$ and 
\begin{equation}
v_\phi=\sqrt{\frac{GM_\star}{r\sin{\theta}}-\xi c_s^2,}
 \label{eq:vphi}
\end{equation}
with $\xi=\alpha+1+f$, here $f=1/2$ is the flaring index (which implies that we consider a constant temperature profile index $\beta\equiv1-2f=0$, that is, a strictly isothermal disc).
The gravitational potential $\Phi$ is given by
\begin{equation}
\Phi=\Phi_\star+\Phi_\mathrm{p},
 \label{eq:potential}
\end{equation}
where
\begin{equation}
\Phi_\star=-\frac{GM_\star}{r},
 \label{eq:Star_potential}
\end{equation}
and
\begin{equation}
\Phi_\mathrm{p}=-\frac{GM_\mathrm{p}}{\sqrt{r'^2+\varepsilon^2}}+\frac{GM_{\mathrm{p}}r\cos\phi\sin\theta}{r_{\mathrm{p}}^2} \, ,
 \label{eq:Planet_potential}
\end{equation}
are the stellar and planetary potentials, respectively. In
Eq.~\eqref{eq:Planet_potential}, $M_\mathrm{p}$ is the planet mass, $r'\equiv\abs{\mathbf{r}-\mathbf{r}_{\mathrm{p}}}$ is
the cell-planet distance, $\phi$ is the azimuth with respect to the
planet, $\theta$ is the colatitude, and $\varepsilon$ is a
softening length used to avoid divergence of the potential in the vicinity of the planet. The second
term on the right-hand side of Eq.~\eqref{eq:Planet_potential} is the
indirect term arising from the reflex motion of the star. Our simulations were
performed with $\varepsilon = 0.01 H_\mathrm{g}$, which is comparable to the cell size
of our computational mesh (see below). We performed additional experiments applying a larger $\varepsilon$ and found similar results
(see Appendix~\ref{sec:appendixB}). It is important to note that, in three dimensions, the parameter
$\varepsilon$ is purely a numerical device and has no physical meaning (if the planet is a point mass).

As done by \citet{MB2016}, our nominal models do not use any tapering function to gradually
introduce the planet mass in the disc. However, we did apply mass tapering in models presented
in Appendix~\ref{sec:appendixB} to demonstrate that our nominal models are not affected
by artificially large vorticity disturbances that might arise in inviscid disc-planet simulations
due to ill-posed initial conditions \citep[see][for an example]{OKS2015}.

The dust density profile is initialized from the gas density using the constant dust-to-gas mass 
ratio $\epsilon\equiv\rho_\mathrm{d}/\rho_\mathrm{g}=0.01$. The initial radial and vertical components of the dust velocity are $u_r=u_\theta=0$, respectively, 
while the azimuthal component is $u_\phi=v_\phi$.

In all our simulations, the planet is held on a fixed non-inclined and circular orbit. 
The dust is sampled by separate fluids with fixed Stokes numbers in the range $[0.01,0.5]$ 
(see Table \ref{tab:simulations}); the equivalent grain size at $r_0=1$, given 
by Equation~(\ref{eq:St}), is kept constant and applied throughout the domain, in each
simulation. Since we consider the aerodynamic dust back-reaction on gas, we performed
simulations using two fluids (gas and one dust species) at a time.


\subsection{Code and mesh domain}
\label{sec:code_mesh_domain}

To numerically solve Equations~\ref{eq:gas_cont} through \ref{eq:dust_mom},
we use the publicly available hydrodynamic
code FARGO3D\footnote{\url{https://github.com/FARGO3D/fargo3d}}
\citep{BLlM2016,BLlKP2019}, with an implementation of the fast orbital advection algorithm \citep{Masset2000}.  
The grid covers the domain $[r_\mathrm{min},r_\mathrm{max}]=[0.48r_\mathrm{p},2.08r_\mathrm{p}]$ in the radial direction and 
$[\theta_\mathrm{min},\theta_\mathrm{max}]=[\frac{\pi}{2}-h_{\mathrm{p}},\frac{\pi}{2}]$ in colatitude (with $h_\mathrm{p}=0.05$, the disc aspect ratio at the planet position),
hence we simulate only one hemisphere of the disc. The azimuthal extent of the domain is
$[\phi_\mathrm{min},\phi_\mathrm{max}]=[-\pi,\pi]$, covering the entire extent of the disc around the star.
The resolution along each respective dimension is
$(N_r,N_{\theta},N_{\phi})=(3200,100,12560)$, leading to cube-shaped cells
$\Delta r/r_\mathrm{p}=\Delta \theta=\Delta \phi=5\times 10^{-4}$ at the position of the planet.
We note that, although only one scale-height $H_\mathrm{g}$ is covered by the vertical extent of the disc,
we resolve such length at the percent level, $H_\mathrm{g}/100$.
Lastly, because our 3D two-fluid simulations have a high resolution and the computational overhead is considerable,
most of our models were run up to a time of $t=20$ orbits (and some up to $t=50$ orbits).

\subsection{Boundary conditions}

We use boundary conditions for each dust species similar to those implemented for the density and velocity components of the gas 
\citep[for details, see][]{ChM2021}. In the radial direction, we use wave-damping boundary zones as in \citet{dVal2006};
the width of the inner and outer damping rings being at $0.15r_p$ and $0.5r_p$, respectively.
The damping time-scale at the edge of each ring amounts to $0.3$ of the local orbital period.
We emphasize that the radial extension of the domain is large enough to avoid any disturbances,
propagating from the disc boundaries toward the horseshoe region.
Any perturbations found in this region are therefore purely a result of the dynamic interaction
among the gas, dust, and the planet. Additionally, we use damping zones only in the radial direction
because their inclusion in the vertical direction would modify the behavior of dust settling 
and may lead to an overly large dust-to-gas mass ratio in the midplane. At the meridional boundaries, in $\theta_\mathrm{min}$: for the gas and dust densities we use an extrapolation of the initial profiles and anti-symmetric boundary conditions for $v_\theta$ and $u_\theta$, respectively. Lastly, since we consider only one hemisphere of the disc, in $\theta_\mathrm{max}$ (that is, at the midplane): we use reflecting
boundary conditions.

\section{Results}
\label{sec:results}

In this section, we present the results of our numerical simulations.
The individual simulations and the adopted parameters are summarized in
Table~\ref{tab:simulations}.

\subsection{Diagnostics}

We focus here on describing the behavior of the vorticity at the midplane ($\theta=\pi/2$) of the gas disc
as well as on analyzing the disturbances generated in the dust density. \citet{MB2016} found that the vortensity
(i.e., the vorticity divided by the gas density) in 3D inviscid and globally isothermal discs is not conserved
as it would be in 2D discs; instead, vortensity perturbations organized in a stripe-shaped patterns can arise
within the horseshoe region of the planet.
Here, we investigate in detail the influence of the gas vorticity on the dust density
distributions when the two-way dust-gas drag force is taken into account.

\begin{figure}

    \includegraphics[width=\columnwidth]{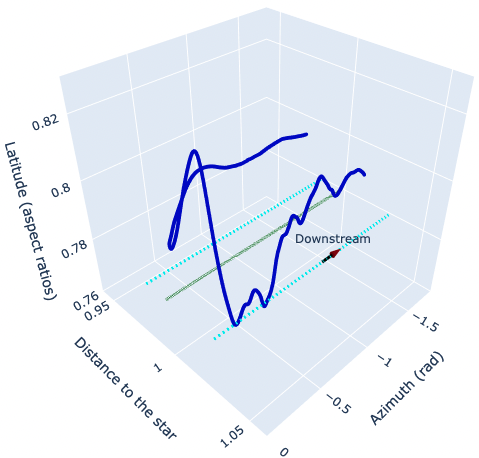}
    \caption{3D-view of a gas streamline in the horseshoe region for the RDISt0Mp3 model.
    The planet is located at 
    $(r,\theta,\phi)=(1,\pi/2,0)$.
    One can see vertical oscillations of fluid elements mainly
    during and after their encounter with the planet (as they move
    downstream, away from the planet), similar to the findings of \citet{MB2016}. 
    The cyan and green dashed lines in the figure represent the boundaries 
    of the horseshoe region (Eq.~\ref{eq:xs}) and the corotation radius,
    respectively.
    }
    \label{fig:stream3d}
\end{figure}

\begin{figure}
  \begin{subfigure}[b]{0.45\textwidth}
  \centering
    \includegraphics[width=\textwidth]{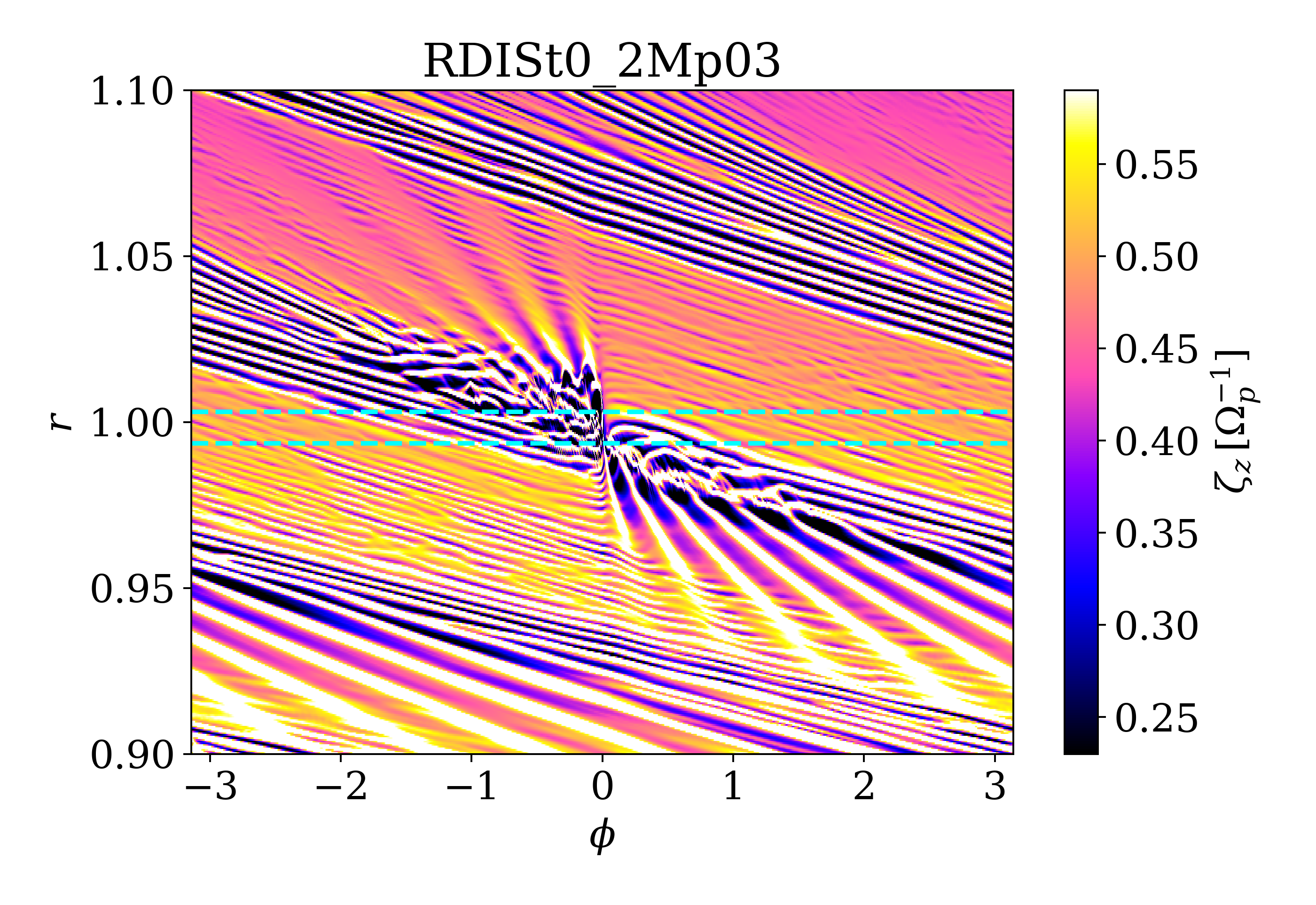}
     
  \end{subfigure}
  \hfill
  \begin{subfigure}[b]{0.45\textwidth}
  \centering
    \includegraphics[width=\textwidth]{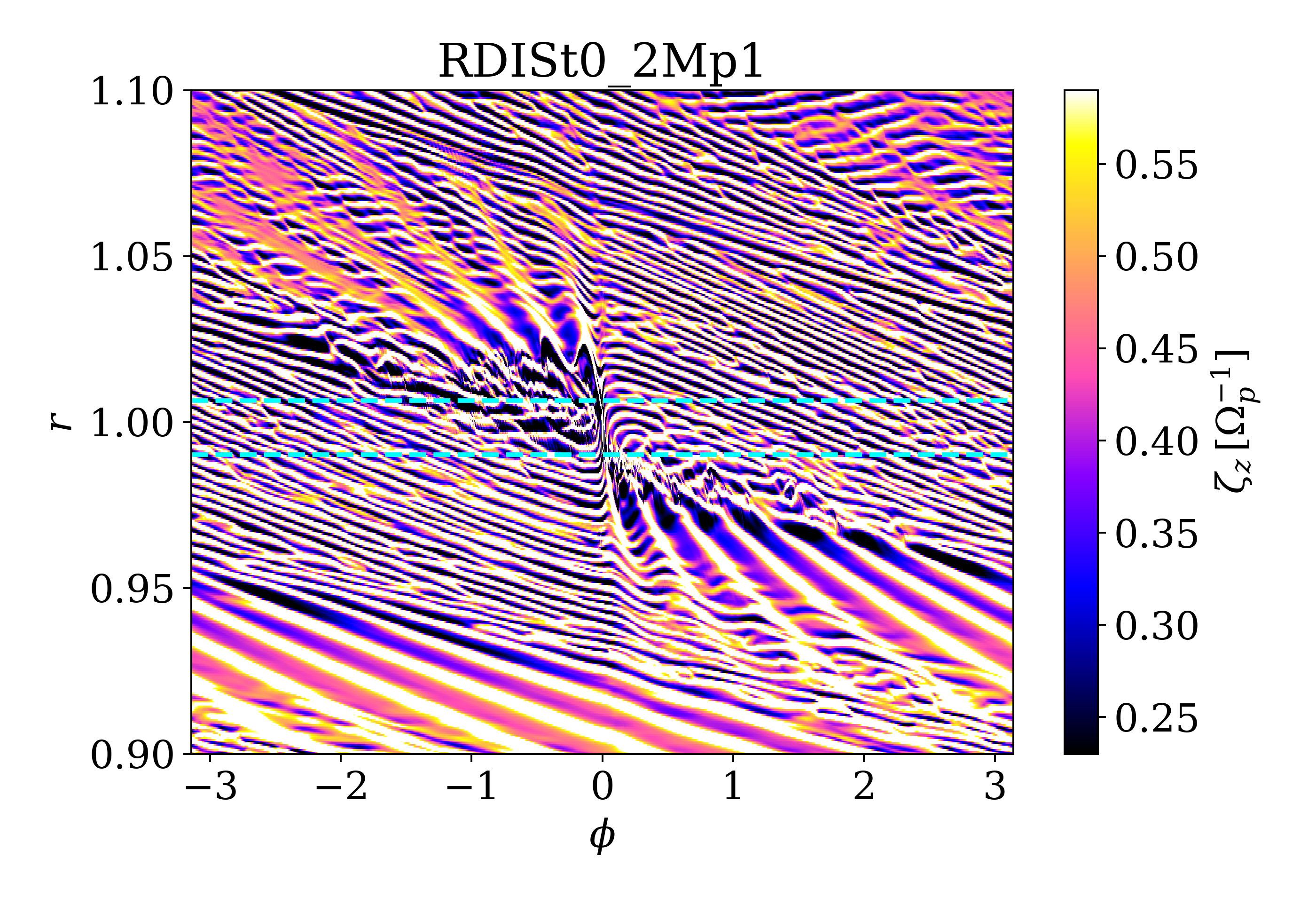}
   
  \end{subfigure}
  \hfill
  \begin{subfigure}[b]{0.45\textwidth}
  \centering
    \includegraphics[width=\textwidth]{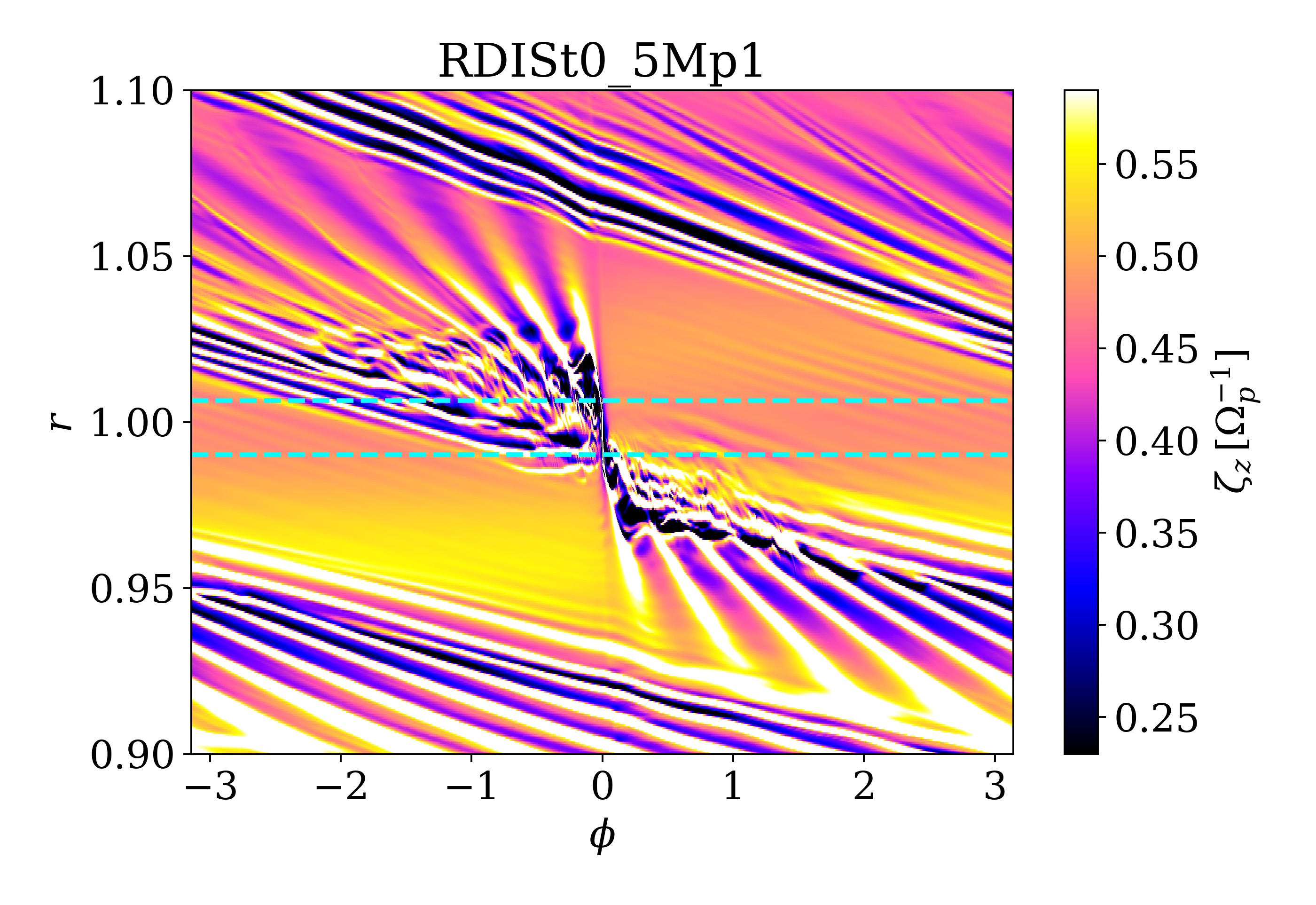}
    
  \end{subfigure}
  \hfill
  \begin{subfigure}[b]{0.45\textwidth}
  \centering
    \includegraphics[width=\textwidth]{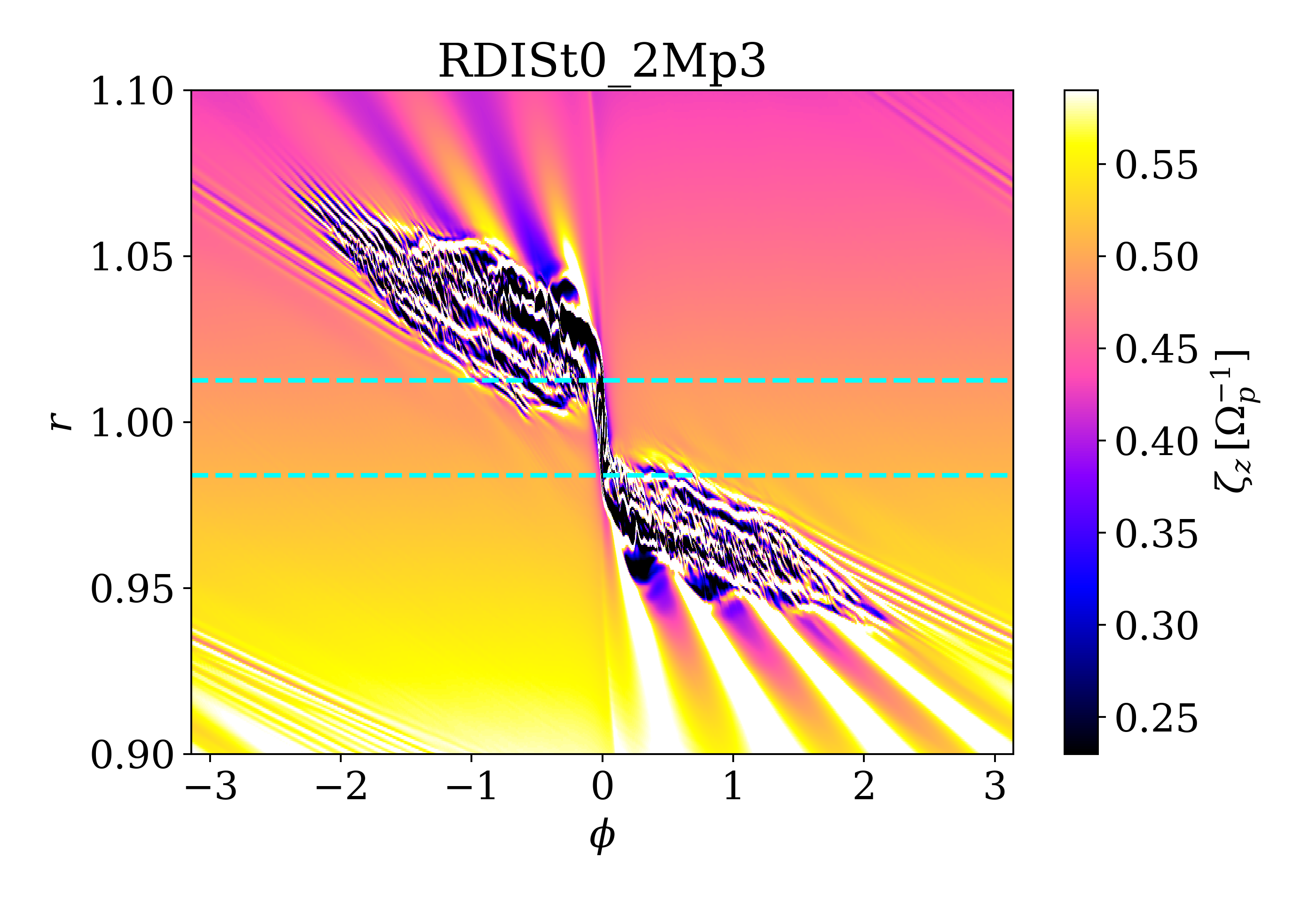}

  \end{subfigure}

  \caption{2$D$-maps of the vorticity at the midplane ($\theta=\pi/2$) 
  for models with large Stokes numbers $St=0.2,0.5$ and different planetary masses 
  (see Table \ref{tab:simulations}) at $t=10$ orbits, 
  except for model RDISt0$\_$2Mp3, which is represented at $t=5$ orbits.
  The cyan dotted lines show the width of the horseshoe region given by Eq.~(\ref{eq:xs}).}
  \label{fig:vortSt020205}
\end{figure}

\begin{figure*}

    \includegraphics[scale=0.45]{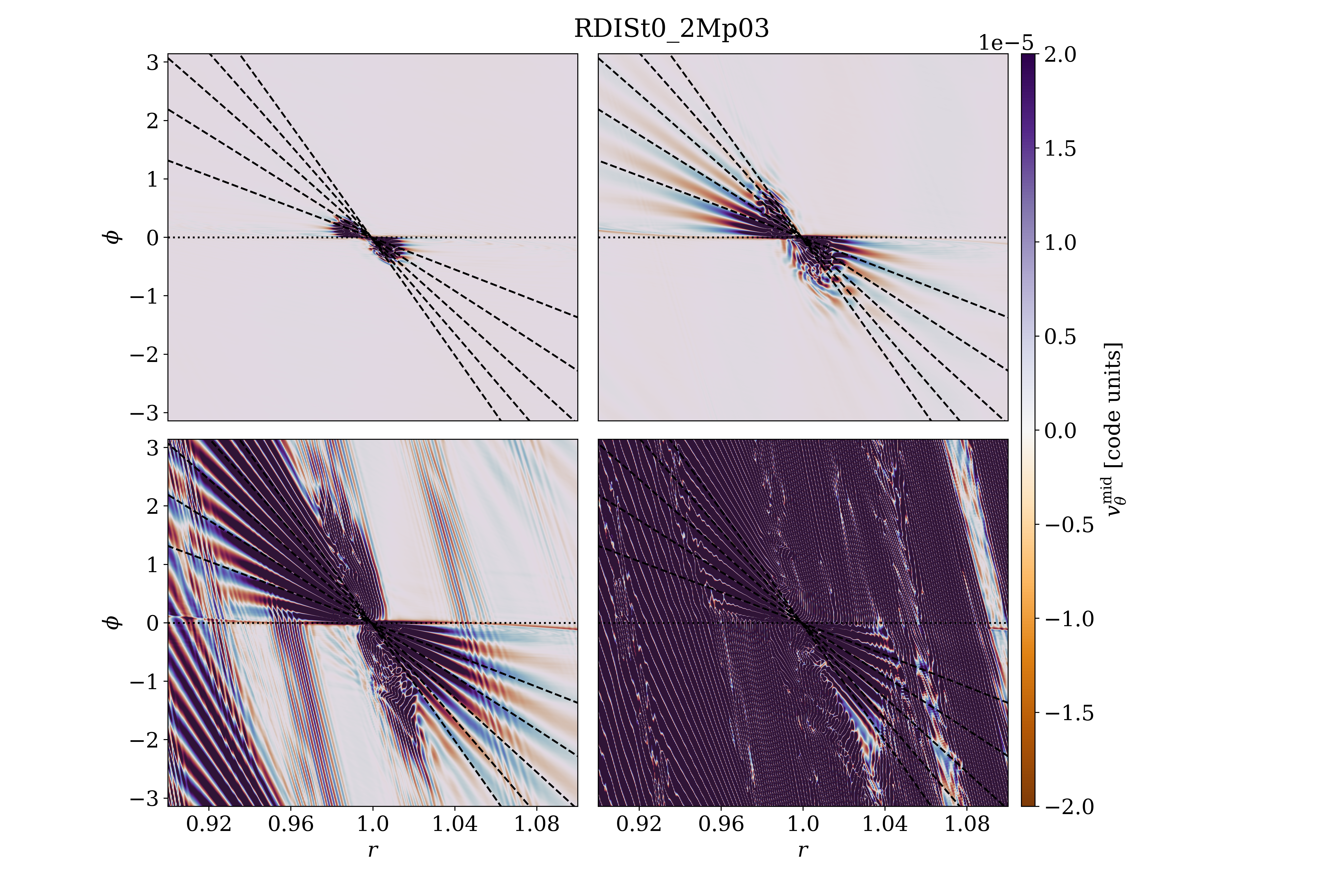}
    \caption{$2D$-maps of the vertical gas velocity component $v_\theta$ (in code units) at the midplane (calculated at $t=5$, $10$, $15$ and $20$ orbits, shown from left to right and top to bottom, respectively) for the RDISt0$\_$2Mp03 model, illustrating the disturbances generated by planetary waves as well as the oscilations (rays) due to the disc bouyancy response as in an adiabatic discs. We overlapped the diagonal dashed black lines which corresponding to the positions of constant phase of linear buoyancy response $\phi_\mathrm{osc}$ given by Eq. (\ref{eq:boyancy}) for an adiabatic disc model in the case when $z=2H_g$ (see text for details).}
    \label{fig:bouyancy}
\end{figure*}

The gas vorticity (in an inertial frame) is calculated as \citep[see][]{ErtelT2004,MB2016}:
\begin{equation}
\bm{\zeta}\equiv\nabla\times\mathbf{v}+2\mathbf{\Omega}_\mathrm{p} \, ,
 \label{eq:vorticity_}
\end{equation}
with $\zeta_z$ the vertical component of the vorticity. 
Following \citet{MB2016}, the horseshoe region is divided in four parts. The first two are
the front and rear regions that contain fluid elements with $\phi>0$ and $\phi<0$, respectively.
Now, each of these two regions can be divided into two parts, called the upstream and downstream
regions. The upstream region is made up of fluid elements
that have not yet experienced a close encounter with the planet while
the downstream region contains all fluid elements that have already experienced a close encounter
with the planet and have crossed the corotation radius $r_c$.
For instance, in Fig.~(\ref{fig:vortMS}), we show the downstream regions (for model RDISt0Mp3)
delimited by $\phi<0$, $r>r_c$ (rear-downstream region) and $\phi>0$, $r<r_c$ (front-downstream region),
respectively. We mention that, as we will show below, it is precisely in these regions that disturbances
in vortensity and dust density originate.

We anticipate that, because there are planetary waves forming near the planet and propagating at the boundaries of the horseshoe region (see Fig. \ref{fig:vortMS}), the steady state condition near the planet does not hold. Furthermore, when the RDI is activated, the gas flow within the horseshoe region becomes eventually chaotic.
Thereafter, the width of the horseshoe region increases considerably.
With the above in mind, we use the definition
of half-width $x_s$ of the horseshoe region, given by \citep[see][]{JM2017}:
\begin{equation}
x_s=\frac{1.05(q/h)^{1/2}+3.4q^{7/3}/h^6}{1+2q^2/h^6}r_p
    \label{eq:xs}
\end{equation}
to delimit this region in all the models presented here. In Eq.~(\ref{eq:xs}), $q\equiv M_\mathrm{p}/M_\star$ is the planet-to-star mass ratio
and $h\equiv H_\mathrm{g}/r$ is the aspect ratio of the gaseous disc. A detailed study on the effects of RDI activation on the width of 
the horseshoe region and on the torque exerted by the planet will be presented in a forthcoming paper (Chametla et al. in preparation).

\begin{figure}
	
    \includegraphics[scale=0.45]{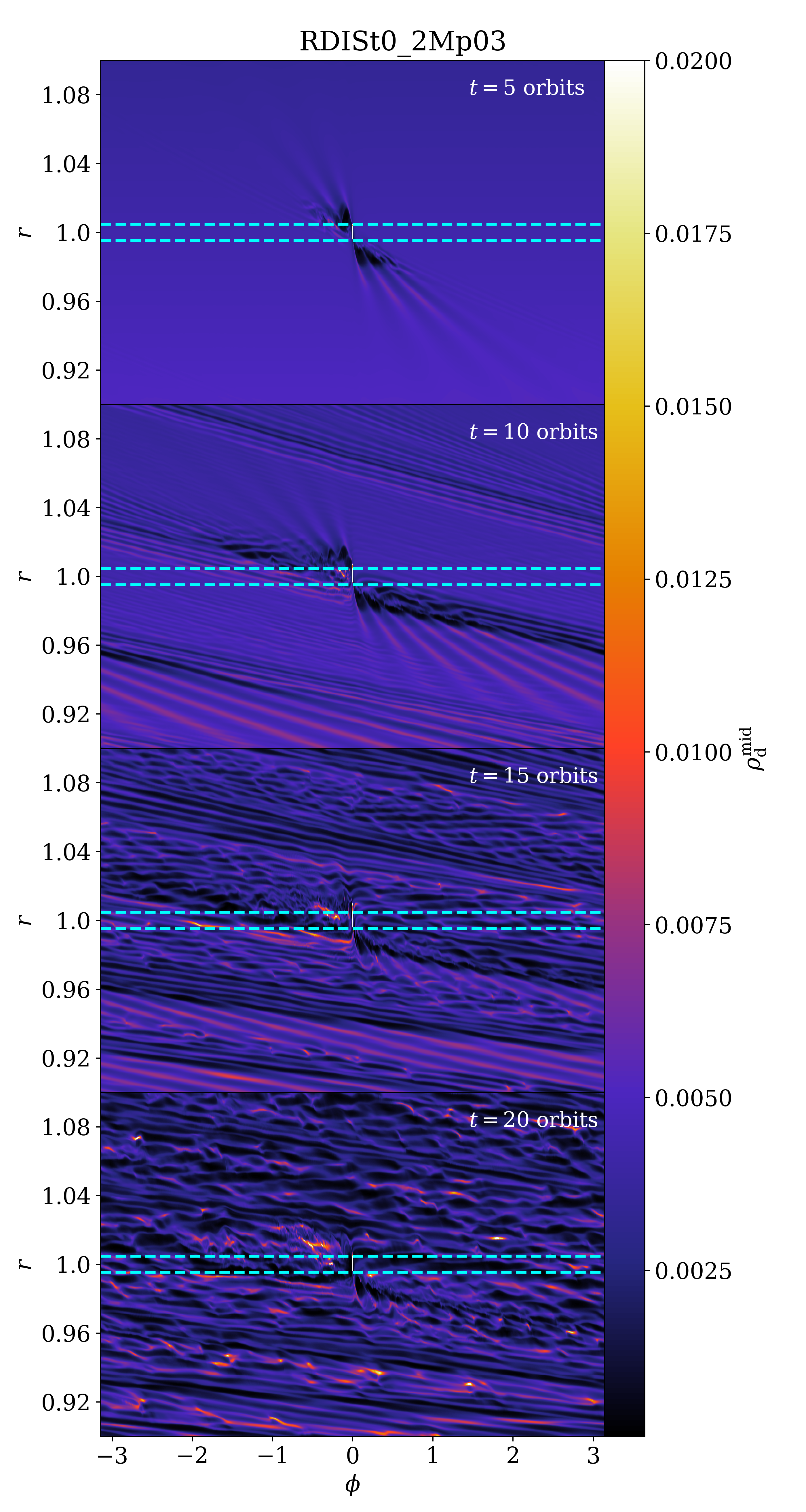}
    \caption{Temporal evolution of the dust density (code units) at the midplane for model RDISt0$\_$2Mp03.
    The cyan dotted lines show the width of the horseshoe region given by Eq (\ref{eq:xs}).}
    \label{fig:RDIdust}
\end{figure}

\begin{figure}
	\includegraphics[width=\columnwidth]{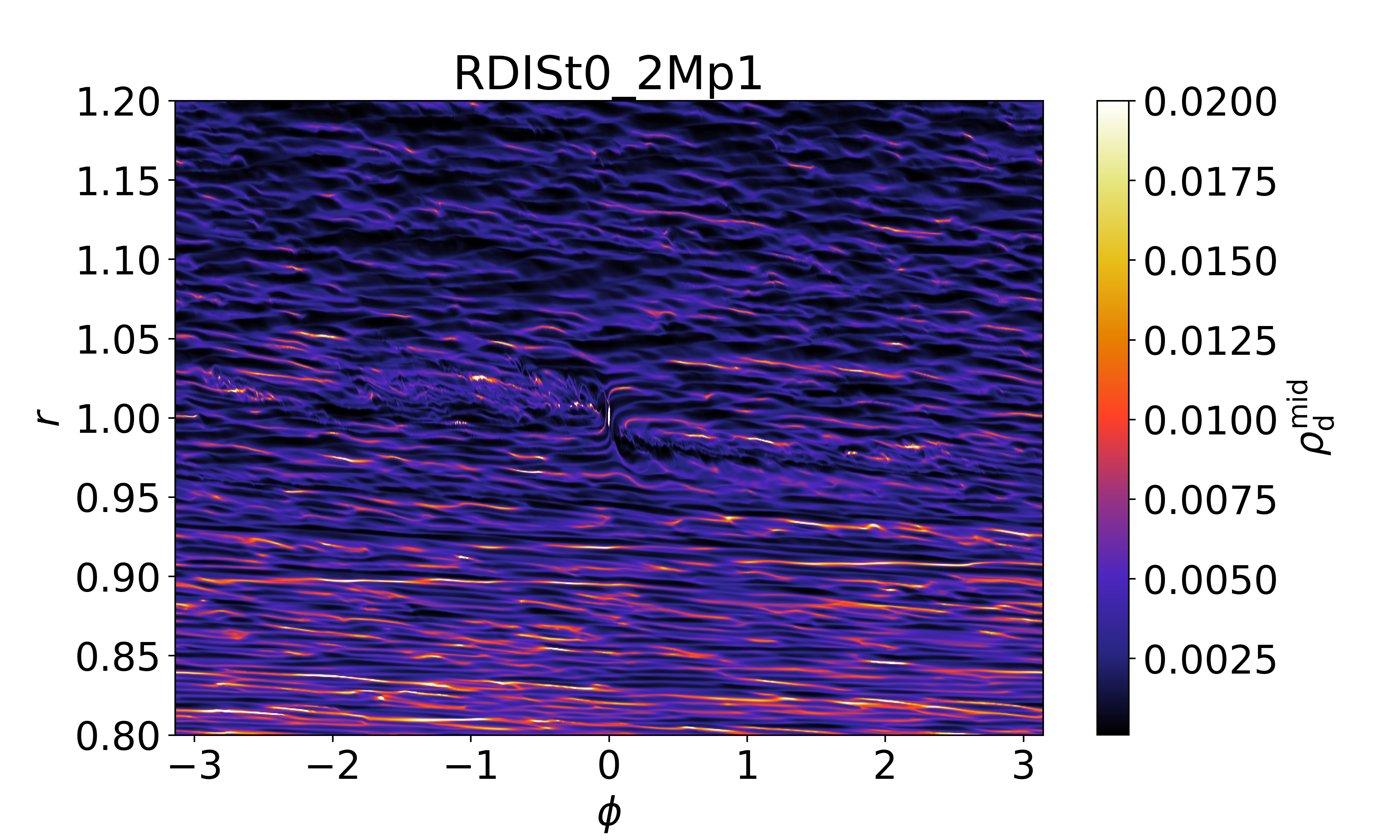}
    \caption{Dust density (code units) in the midplane for the RDISt0$\_$2Mp1 model,
    at $t=20$ orbits. At this time the RDI has propagated beyond the horseshoe region.}
    \label{fig:rdi}
\end{figure}

\begin{figure*}
  \begin{subfigure}[b]{0.49\textwidth}
    \includegraphics[scale=0.45]{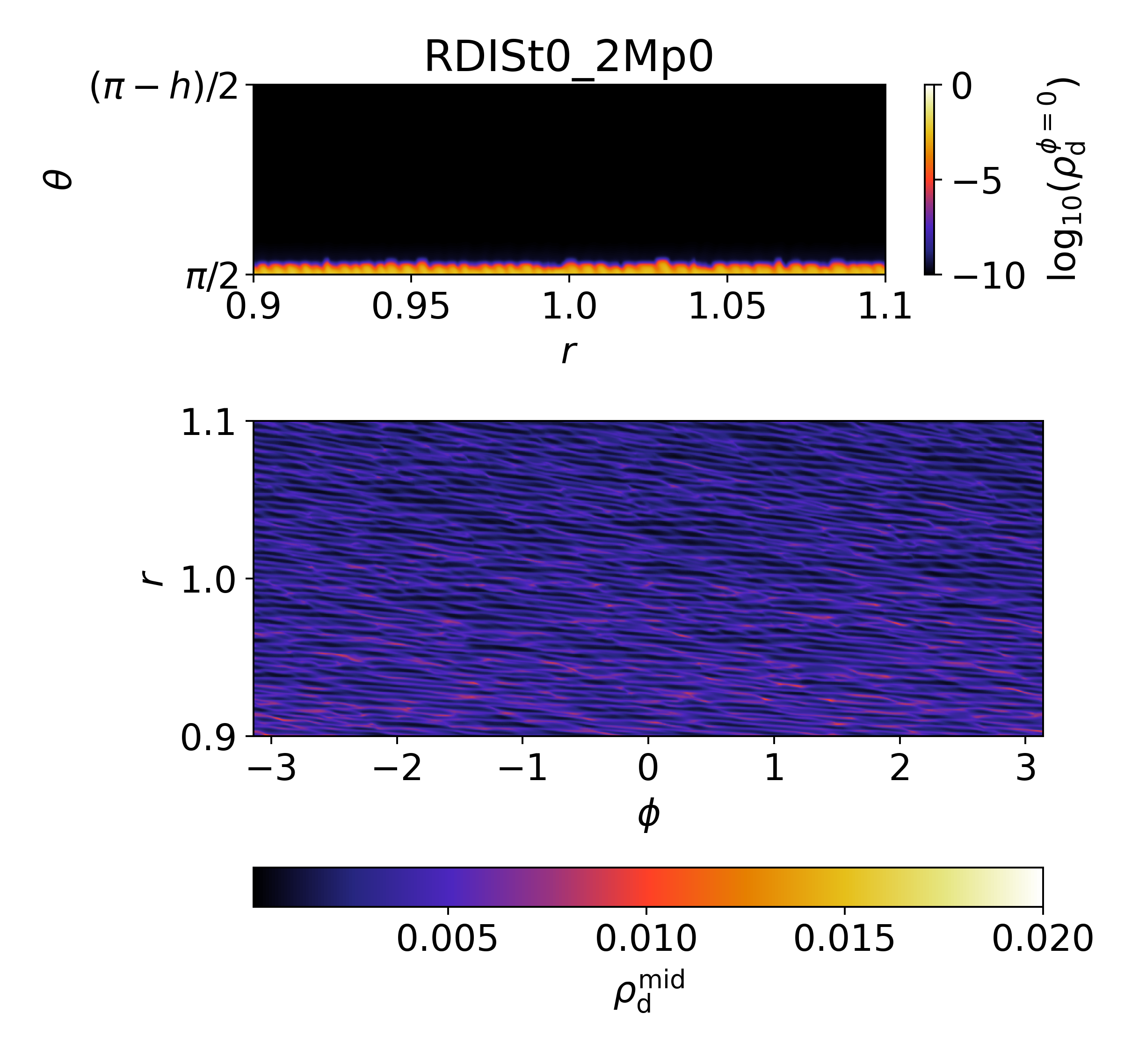}
    \label{fig:f1}
  \end{subfigure}
  \hfill
  \begin{subfigure}[b]{0.49\textwidth}
    \includegraphics[scale=0.45]{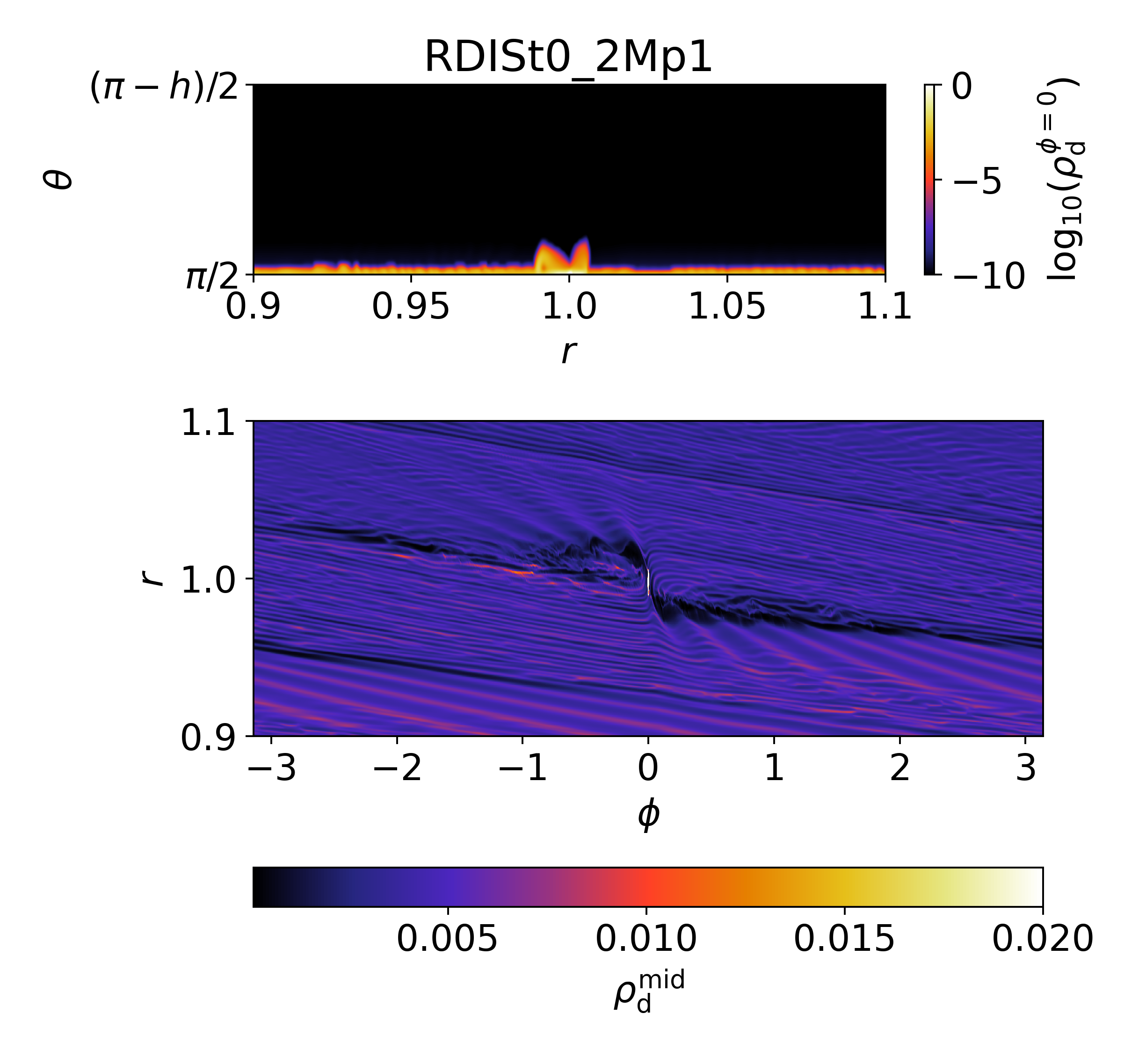}
    \label{fig:f2}
  \end{subfigure}
  \caption{Comparison of the disturbances generated in the dust density (code units),
  after $t=10$ orbits, for a case
  without (model RDISt0$\_$2Mp0, left) and for a case
  with a planet embedded in the disc (model RDISt0$\_$2Mp1, right).}
  \label{fig:settling}
\end{figure*}

\subsection{Planetary waves excited by a planet in a purely gaseous disc}
\label{subsec:planwaves}

Stripe-like substructures found by \citet{MB2016} in the vortensity distribution
within the horseshoe region are caused by vertical oscillations of differential gas elements
as they make a U-turn in front of (or behind) the planet.
We verified that such vertical oscillations persist in all our dust-free high-resolution
simulations. In Fig.~\ref{fig:stream3d}, we show a three-dimensional view of a gas streamline for the RDISt0Mp3 model. It can be seen that the streamline propagating above the midplane exhibit oscillations close to the planet, which produces the stripes in the midplane of the disc in the vicinity of the planet. Material compression towards the midplane decreases the midplane
vortensity, while expansion increases it.
Nevertheless, it is important to emphasize that these vorticity perturbations
alone cannot produce filamentary concentrations of dust since we never detect
such concentrations in simulations in which the aerodynamic dust feedback is neglected 
(see Section~\ref{subsection:dust_density} and Appendix~\ref{sec:appendixC}).

Planetary waves generated in the vicinity of a planet, and propagating in the horseshoe region in an inviscid gas disc (not including dust), can be observed 
even in 2D high resolution simulations \citep[see][]{KLL2003,Li_etal2005}. Here, we provide a detailed analysis 
of the formation and propagation of planetary waves in the region near the planet. 
To confirm that there are indeed planetary waves propagating from the vicinity of the Hill radius, 
Fig.~\ref{fig:vortMS} shows the gas vorticity at the midplane ($\theta=\pi/2$) for the dust-free RDISt0Mp3 model,
calculated at $t=20$ orbits. One can clearly see the propagation of planetary waves from the vicinity of the planet,
located at $(r_\mathrm{p},\phi_\mathrm{p})=(1,0)$. 

The planetary waves are predominantly propagating downstream in the horseshoe region, along its critical streamlines. The disturbances in the vorticity at $t=20$ orbits for the RDISt0Mp3 model can by directly compared with the results of \citet[][see their Figure~9]{MB2016}.
Such a comparison shows that the stripes they reported are also present (and sharper) in our simulation.
Remarkably, small vortices that appear in our simulation were not present in \citet{MB2016}.
We attribute these differences to the higher mesh resolution used in our study.

It should be noted that the planetary waves propagate along the downstream regions close to the separatrices
while the stripes remain within the horseshoe region (see Fig.~\ref{fig:vortMS}).
When the feedback of the dust on the gas is not present, there is no indication
of ``interaction'' between the stripes and the planetary waves.
Lastly, we mention that in dust-free models RDISt0Mp03 and RDISt0Mp1, we find that the vorticity perturbation
generally weakens and the region of propagation of PWs shrinks as the perturbing mass decreases.
The latter is to be expected since PWs propagate at the edges of the horseshoe region whose width
is proportional to the square root of the planet mass (see Eq.~\ref{eq:xs}).

\subsection{Enhanced propagation of planetary waves in dusty gaseous discs}\label{subsection:EPP}

In simulations that account for different dust species (that is, $St\neq0$) and their
feedback on the gas, the gas vorticity distribution evolves differently compared to the models without dust
($St=0$). For instance, a direct comparison between the model RDISt0Mp3 (see Fig.~\ref{fig:vortMS}) and model
RDISt0$\_$2Mp3 (see Fig.~\ref{fig:vortSt020205}) shows that the disturbances spread faster in the case that
accounts for the feedback. In fact, for model RDISt0$\_$2Mp3, barely 5 orbital periods have elapsed.

In Fig.~\ref{fig:vortSt020205}, we also show the gas vorticity at the midplane for the RDISt0$\_$2Mp03 and RDISt0$\_$2Mp1 models, calculated at $t=10$ orbits.
In these two models, the Stokes number is fixed ($St=0.2$) and we compare two different masses of the planet ($M_\mathrm{p}=0.3M_\oplus$ and $M_\mathrm{p}=1M_\oplus$, respectively).
One can see that the vorticity perturbations extends over a larger and larger region, as the planet mass increases.
Once can also see that sets of colored-rays (blue-white-yellow rays) and black-white stripes appear 
(the latter are named pw-stripes hereafter, see Appendix \ref{sec:appendixA}), which propagate from the edges
of the horseshoe region. In the RDISt0$\_$2Mp1 model the pw-stripes propagate somewhat faster, 
since in the same amount of time they can be seen to have fully spread into and out of the horseshoe region,
overlapping with the colored-rays.
We emphasize that while the colored-rays maintain a marginal inclination with respect to the radial axis
(i.e., the direction perpendicular to the horseshoe region), whereas the pw-stripes exhibit a greater inclination, which causes them to also propagate in the upstream regions.

\subsection{Buoyancy resonances in a dusty, strictly isothermal gaseous disc}
\label{subsec:br}

To understand the differences in the vorticity maps of purely gaseous and dusty gaseous discs, we analyse the buoyancy response of the disc to the planet for the case of a dusty gaseous disc. 
This analysis has been presented in other studies for the case of adiabatic gaseous discs without dust 
\citep[see][]{Zhu_etal2015,MC2020,Han2022,Ziampras2023}, therefore, we will provide a brief explanation which justifies its application
to a strictly isothermal disc.

In an adiabatic disc (not including dust), the fluid elements approach to the planet, along the azimuthal direction, at a given height above 
the disc midplane but, due to the planet’s gravitational potential, they experience vertical oscillations around their equilibrium positions with a frequency given by the Brunt–V\"ais\"al\"a frequency,
\begin{equation}
N(z)=\sqrt{\frac{g_z}{\gamma}\frac{\partial}{\partial}\left[\ln{\left(\frac{P}{\rho^\gamma}\right)}\right]},
    \label{eq:Brunt-Vf}
\end{equation}
with
\begin{equation}
g_z=-\frac{\partial \Phi_\star}{\partial z}.
    \label{eq:gz}
\end{equation}
Therefore, if we assume that the zero phase is at the planet position $\phi=\phi_\mathrm{p}=0$, the vertical oscillations due to the buoyancy response are given by lines of constant phase at azimuth:
\begin{equation}
\phi_\mathrm{osc}=-2n\pi\sqrt{\frac{\gamma}{\gamma-1}}\left(\frac{\Omega_p-\Omega}{\Omega_\mathrm{K}}\right)\left(\frac{H_\mathrm{g}}{z}\right)\left(1+\frac{z^2}{R^2}\right)^{3/2},
    \label{eq:boyancy}
\end{equation}
where $\Omega_\mathrm{K}(R)=\sqrt{GM_\star/R^3}$ is the Keplerian angular velocity, $n$ is an integer and $R$ and $z$ are the radial and vertical cylindrical coordinates, respectively \citep[see][]{Zhu_etal2015}. 

As typical in this type of analysis, we show in Fig.~\ref{fig:bouyancy} $2D$-maps of the vertical component 
of the velocity $v_\theta$ at the disc midplane (equivalent to a height $z=0$ in cylindrical coordinates)
at four different times ($t=5,10,15$ and $t=20$ orbits, shown from left to right and top to bottom in the figure, respectively)
for the RDISt0$\_$2Mp03 model. At $t=5$ orbits, we can see the formation of perturbations in the vertical velocity map along the downstream flow
of the planet. At $t=10$ orbits, it is possible to identify two different types of disturbances; the first is a set of rays (colored rays)
propagating beyond the corotation region with a well-defined pattern, which resembles the pattern resulting from buoyancy resonances.
We encourage the reader to compare this figure with Figs.~12, 2, and 5 presented, respectively, in \citet{MC2020}, \cite{Han2022} 
and \citet{Ziampras2023}.

To show that in fact these rays have the same behavior as the rays formed by the buoyancy response,
on Fig.~\ref{fig:bouyancy} we superimpose the predicted pattern given by Eq.~(\ref{eq:boyancy})
for the case of an adiabatic disc, considering that $z=2H_\mathrm{g}$\footnote{The perturbations in the vertical component $v_\theta$ of the velocity, associated with bouyancy resonances do not appear for values of $z$ at which the dust has already settled.}.
We find that the positions of these rays match the predicted pattern for the buoyancy resonances from 
Eq.~(\ref{eq:boyancy}). The second type of perturbations that can be identified in Fig.~\ref{fig:bouyancy} 
is located exactly in the positions where the planetary waves are launched. 
At a time $t=15$ orbits, we can see that the second type of perturbations is propagating in a different direction 
than that generated by the bouyancy resonances. We can thus conclude that this perturbation is the same one
we identified in Section~\ref{subsection:EPP} as pw-stripes in the vorticity maps 
(see Fig.~\ref{fig:vortSt020205}). We emphasize that while the bouyancy resonance rays maintain
their orderly propagation, the pw-stripes are propagating beyond the corotation region
and in places where the bouyancy resonance rays do not develop. Lastly, at a time $t=20$ orbits, the pw-stripes have almost completely populated the vertical velocity $2D$-map inside
and outside the corotation region, overlapping the rays of the bouyancy resonances, and 
emerging as the dominant perturbations.

In summary, bouyancy resonances emerge in our dusty, strictly-isothermal gaseous disc models.
This outcome appears in accord with recent studies on isothermal dusty gaseous discs and 
adiabatic purely gaseous discs \citep[see, e.g.,][]{LY2017,Lin2019}.

\subsection{Relationship between bouyancy resonance rays and planetary waves}
\label{subsec:relationship}

In general, in all our numerical models where we include the drag force of the dust on the gas, 
we find that the vorticity perturbations are significantly different from the vorticity perturbations
in purely gaseous disc models (i.e., $St=0$ models). Clearly, the explanation for the differences
in these vorticity disturbances lies in the feedback effect of the dust on the gas.
However, it must be kept in mind that when a planet is embedded in a gas disc (not including dust),
it produces two different types of vorticity disturbances in the horseshoe region 
(as we discussed in Section~\ref{subsec:planwaves}). 
For the first type, the so-called stripes \citep[see][]{MB2016}, the vorticity disturbances
are generated by vertical excursions of the gas fluid elements as they pass close to the planet 
during the execution of its U-turn. 
For the second type, the planetary waves, the disturbances result from a vorticity gradient
in the vicinity of the planet.

In dusty discs, the patterns found in the vorticity distribution emerging from planetary waves show 
additional characteristics (see Fig.~\ref{fig:vortSt020205}). For instance, there is a greater contrast
in the vorticity maps that go from black to white, as well as a different inclination with respect 
to the bouyancy resonance rays, as discussed above.

We note that, although planetary waves could produce a boost in the propagation of the bouyancy resonance rays
beyond the edges of the horseshoe region, the resulting pattern from this disturbance in the vorticity can be distinguished from respect to the pattern generated by the pw-stripes.
Such pattern holds for different Stokes numbers and for different planetary masses.
For example, let us qualitatively compare the vortensity perturbations of the models 
RDISt0$\_$2Mp03, RDISt0$\_$2Mp1 and RDISt0$\_$5Mp1 models of Fig.~\ref{fig:vortSt020205},
which are represented at the same orbital time. We mention that, the number of well-defined rays in the upstream and downstream regions
are similar among all models. In addition, we verified that the inclination of each of these stripes relative to the radial axis (i.e., vertical axis in the plane $r-\phi$)
is basically the same in both regions (upstream and downstream).
Lastly, we checked that the intensity of these disturbances is similar in the three models.
As a consequence of the similarities among these bouyancy resonance rays, we argue that
(by themselves) they cannot generate all the other disturbances observed inside and outside
the horseshoe region. In other words, if these rays were the cause of all vorticity perturbations, the black and white contrast should populate the upstream regions in the horseshoe region and beyond in all three models, which does not occur.
We finally note that, by including the two-way drag force between dust and gas,
the disturbances generated in the gas can be seen reflected in the dust dynamics,
as discussed below.

\subsection{Streaming Instability-like perturbations arising in the dust density}
\label{subsection:dust_density}

So far we showed that a planet embedded in a gas disc can generate planetary waves propagating at the edges of the horseshoe region.
In addition, we showed that the development of these perturbations is enhanced by the aerodynamic feedback of dust drag
onto the gas. As a next step, we study the evolution of the dust density, focusing
on the model RDISt0$\_$2Mp03 (Fig.~\ref{fig:RDIdust}).

We uncover global-scale perturbations of the dust density that we attribute to the RDI
and its activation by the PWs. Clearly, the structures developing in $\rho_{\mathrm{d}}$
can be associated with those found for the vorticity. After a few orbits, 
the RDI in the dust density propagates over the region affected by the vorticity disturbances.
In support this statement,
one can compare the dust density after $10$ orbits with the vortensity shown in the upper panel
of Fig.~\ref{fig:vortSt020205}. 
At later times, a filamentary structure in $\rho_{\mathrm{d}}$ appears and extends well
beyond the radial width of the horseshoe region. To emphasize this finding, we show
a radially larger portion of the disc in Fig.~\ref{fig:rdi} for model RDISt0$\_2$Mp1,
at $t=20$ orbits.

Although the pattern in the dust density resembles the SI, here it arises from the similarity
between the propagation velocity of gas-vorticity waves caused by the planet and the dust-gas
relative velocity components.
To demonstrate the differences between the classical SI and the planet-induced RDI,
Fig.~\ref{fig:settling} shows snapshots of the dust density at the midplane ($\theta=\pi/2$)
and in the vertical plane ($\phi=0$) for model RDISt0$\_$2Mp0 (left) and model RDISt0$\_$2Mp1
(right), at $t=10$ orbits. For these models, which have dust with the Stokes number $St=0.2$,
we find a fast settling of dust towards the midplane of the disc (as expected since
the dust is loosely coupled to gas and there is no vertical stirring mechanism
in our inviscid disc). In the planet-free simulation, RDISt0$\_$2Mp0, we included 
a loud white noise in the radial and vertical components of the gas and dust velocities,
with an amplitude $10^{-2}c_s$, which favors a rapid development of dust instabilities
\citep[see for instance][]{Krapp_etal2020}.

By means of a direct comparison of the two models without and with a low-mass planet embedded
in the disc (see Fig.~\ref{fig:settling}, left and right panels, respectively), it becomes clear
that the fast settling of the dust, and the related increase of the dust-to-gas mass ratio
in the midplane, in model RDISt0$\_$2Mp0 does not trigger other instabilities beyond the SI.
However, when there is a low-mass planet embedded in the disc (see Fig.~\ref{fig:settling}, right),
strong disturbances in the dust density appear, which arise from the position of the planet
and propagate along the edges of the horseshoe region.
Additionally, some disturbances can also be seen in the vertical direction
(in the $r-\theta$ plane), which we interpret as outcomes of weak planet-induced
vertical stirring.
Following \citet{SH2018}, we also consider the fact that it is possible for several RDIs 
to act simultaneously (for example, SI alongside the dust settling instability,
among others). However, it is important to note that the RDI instability reported
here may have a potential impact on dust dynamics both in the horseshoe region and
in regions beyond it, because of the effects of pw-stripes on 
the gas vortensity when dust feedback is taken into account. In fact, as we show in Appendix~A, the formation of a single small planetary
wave in the gas disc produces pw-stripes in the dust that can reach a considerable
radial and azimuthal extent in a short dynamical time (see Fig.~\ref{fig:pws}).

\section{Discussion}
\label{sec:discussion}

\subsection{Role of Dust Settling}
\label{subsec:DSI}

Since in our models we have considered a strictly isothermal disc as well as Stokes numbers $St\geq0.01$ and included the feedback of dust onto the gas, we do not expect either vertical shear instability (VSI) \citep[see for instance][]{Lin2019} or dust settling instability \citep[DSI;][]{SH2018,SH2020Physicalmodels,Lin2019,Krapp_etal2020} to develop and, therefore, the dust should settle in a thin layer in the midplane of the disc. In fact, in Fig. \ref{fig:settling}, for a Stokes number $St=0.2$, one can clearly see how the dust accumulates in a thin layer within the first 10 orbits of the planet. The latter, while producing stirring in its vicinity due to its mass, is not strong enough to increase the height scale of the dust beyond $0.2h_\mathrm{g}$.

\begin{figure}
  \begin{subfigure}[b]{0.49\textwidth}
  \centering
    \includegraphics[width=\textwidth]{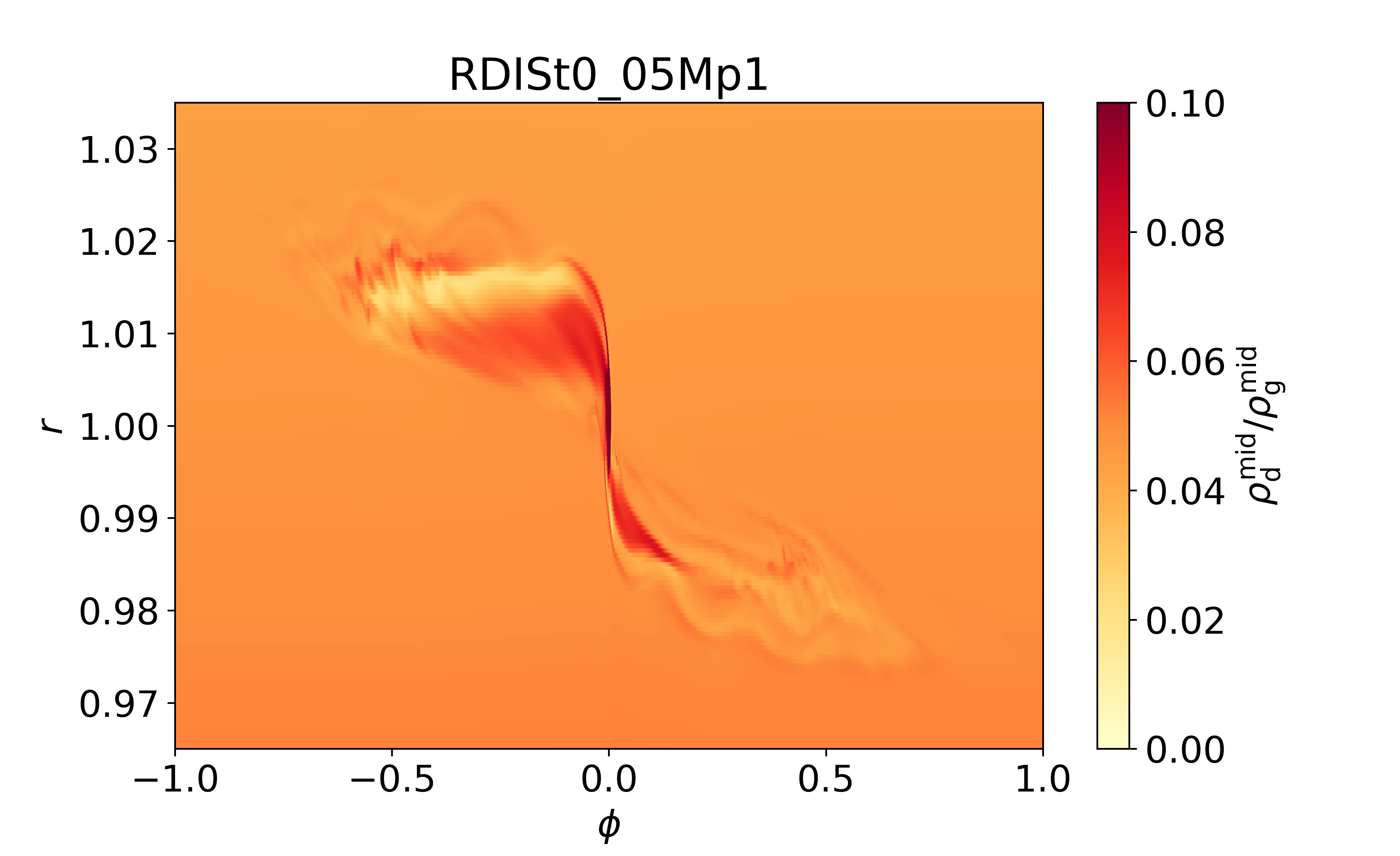}
     
  \end{subfigure}
  \hfill
  \begin{subfigure}[b]{0.49\textwidth}
  \centering
    \includegraphics[width=\textwidth]{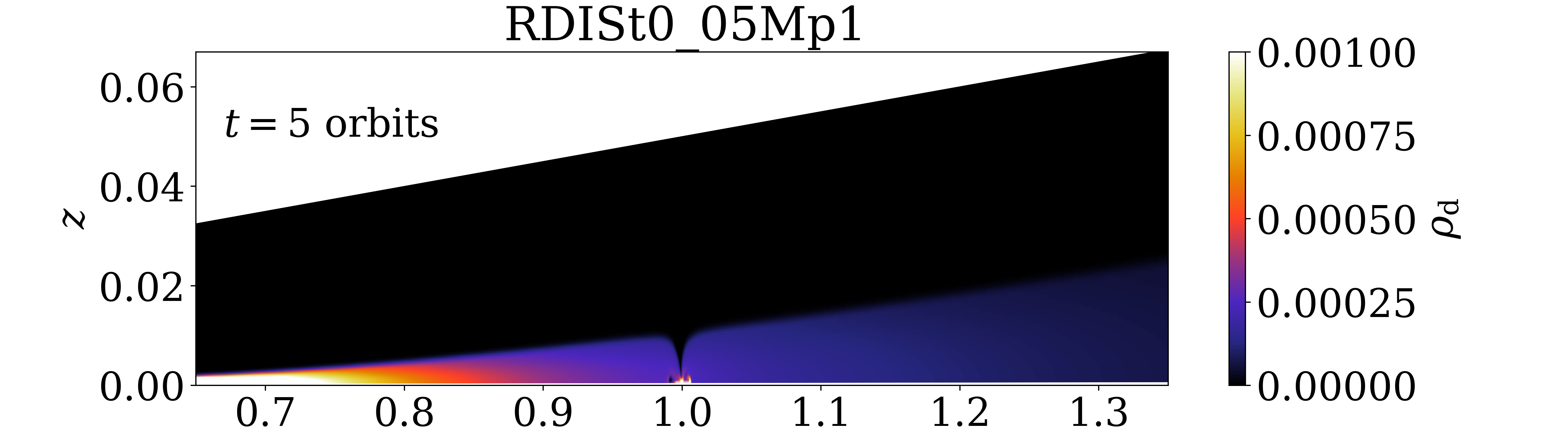}
   
  \end{subfigure}

  \caption{\textit{Top}. Dust-to-gas density ratio, $\epsilon$, at the disc midplane for model RDISt0$\_$05Mp1 at $t=5$ orbits.
  \textit{Bottom.} Vertical dust density distribution (code units) for the RDISt0$\_$05Mp1 model.}
  \label{fig:dgratio}
\end{figure}

In the case of a lower Stokes number, as in the RDISt0$\_$05Mp1 model, we find that the dust is concentrated in a layer not exceeding $0.25h_\mathrm{g}$ (see lower panel in Fig. \ref{fig:dgratio}) within the first five orbital periods of the planet. Interestingly, dust fall is observed at the planet's pole, and there is only very weak stirring generated by the planet, resulting in the dust being confined within this thin layer. As a consequence of this, we again find no development of the VSI or DSI. Note that, the dust-to-gas mass ratio remains small and it does not contribute to the formation of substructures beyond the horseshoe region. In other words, there is no indication that increasing the gas-dust ratio (which remains much lower than unity) triggers SI. Therefore, it is clear that the settling of dust
by itself does not generate an instability of the type reported here. Nevertheless, as the dust exerts a drag force on the moving gas, 
rapid settling of dust could be enhanced outside the horseshoe region, where the stripes
form in the vortensity of the gas (see Fig.~\ref{fig:vortSt020205}).

\subsection{Growth rate of PW resonant drag instability as function of Stokes number, dust-to-gas mass ratio and planet mass}
\label{subsec:dSt}

Our results suggest that there is a slight implicit dependence of the RDI on the Stokes number 
via the relative fast or slow settling. As we discussed in Section~\ref{sec:results}, for the models
that include a planet of mass $M_\mathrm{p}=1 M_\oplus$ and different Stokes numbers,
disturbances in the vorticity of the gas become more intense as the Stokes number increases,
but they propagate faster for intermediate Stokes numbers (see Fig.~\ref{fig:vortSt020205}).
The first occurrence can be explained as follows. If dust settling is fast, then a gas
parcel located high above the disc midplane only feels a drag force from the dust for
a very short time. Conversely, when a gas parcel is very close to the midplane,
it should increase its oscillations because it feels a greater drag force from
the dust, which has settled from different heights. Otherwise, as the settling time increases, the feedback produces longer oscillations.

We find that the formation of the PW resonant drag instability does not depend on
the dust-to-gas mass ratio, $\epsilon$. The only influence of this parameter on
the RDI regards the propagation of the instability. As a demonstration that
the activation of the RDI does not require that $\epsilon>1$, in Fig.~\ref{fig:dgratio}
we show dust-to-gas density ratio at the midplane of model RDISt0$\_$05Mp1.
We can see that $\epsilon\ll 1$ in the most parts where the disturbances propagate.
This result has important consequences, since it means that the dust instability
develops even when dust-to-gas density ratios are relatively small. 

Finally, we mention that, while the PW resonant drag instability is driven by the planet potential and gas-dust interactions, the propagation of this instability does not significantly depend on the mass of the planet (see, for instance Figs.~\ref{fig:RDIdust} and \ref{fig:rdi}).

\begin{figure}
  \begin{subfigure}[b]{0.49\textwidth}
  \centering
    \includegraphics[width=\textwidth]{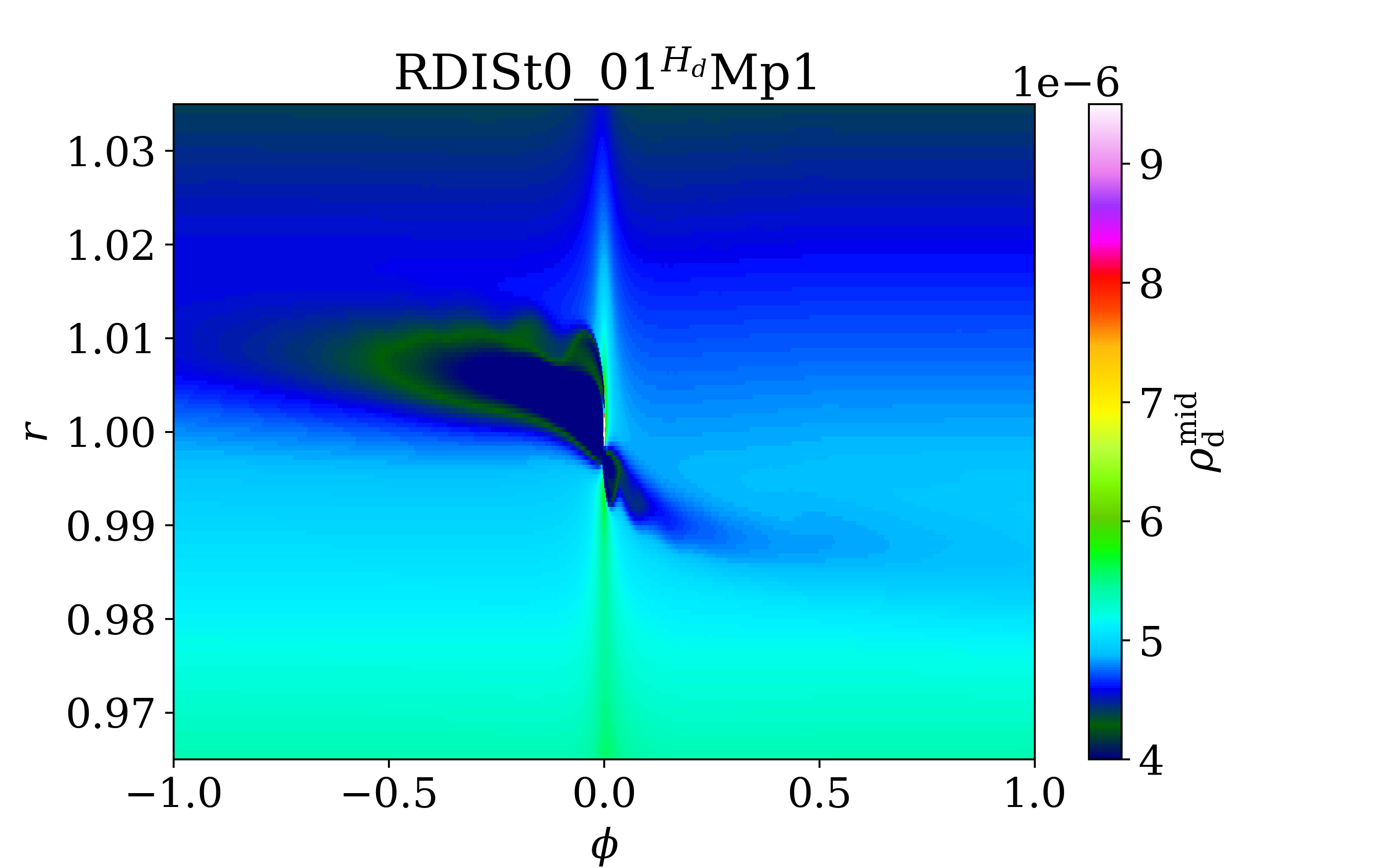}
     
  \end{subfigure}
  \hfill
  \begin{subfigure}[b]{0.49\textwidth}
  \centering
    \includegraphics[width=\textwidth]{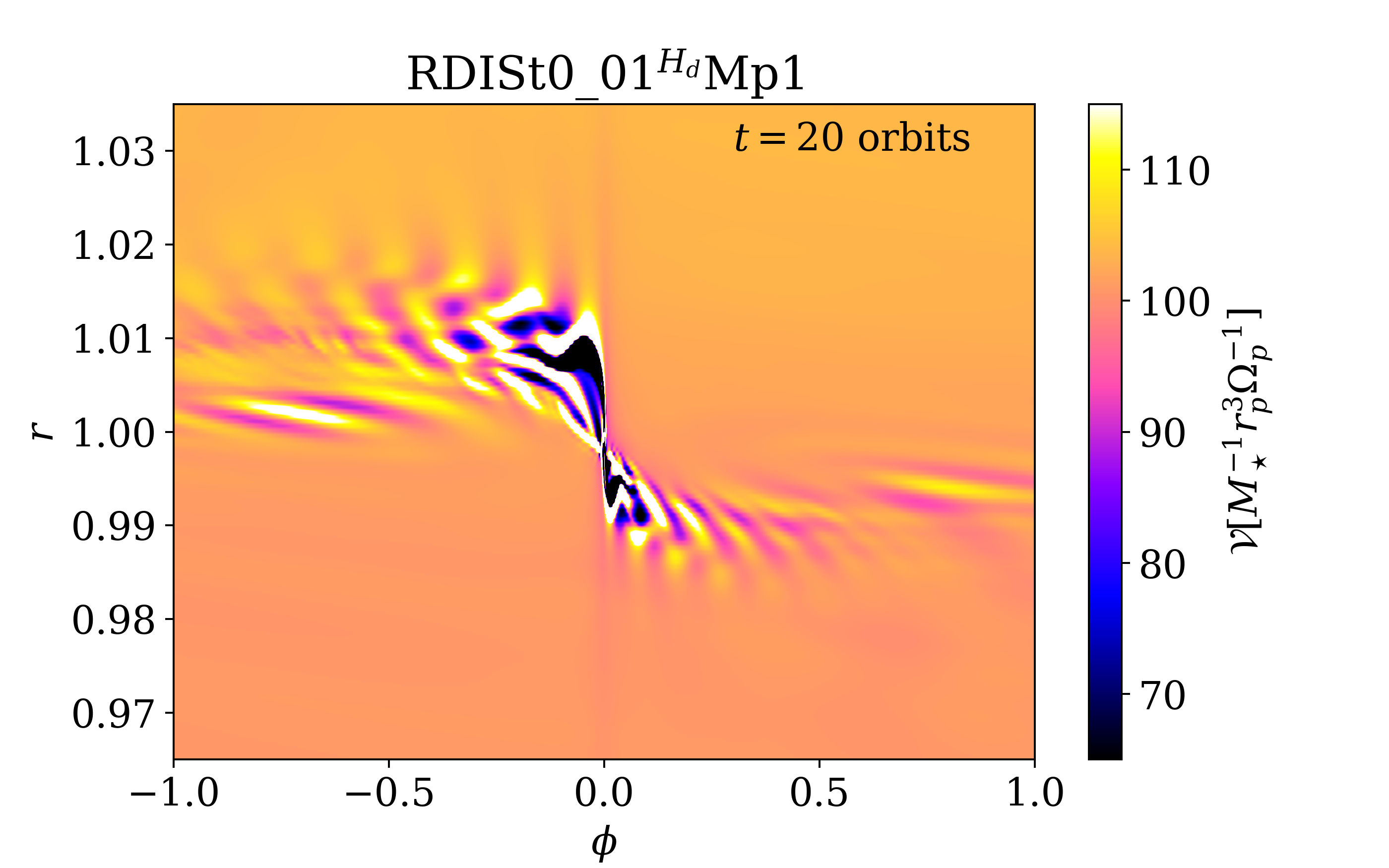}
   
  \end{subfigure}

  \caption{\textit{Top}. Dust density distribution (code units) in model RDISt0$\_$01$^{H_d}$Mp1,
  calculated at the midplane, at $t=20$ orbits.
  \textit{Bottom.} Vortensity $(\mathcal{V}\equiv\zeta_z/\rho_\mathrm{g})$ at the midplane
  in model RDISt0$\_$01$^{H_d}$Mp1, calculated at the same time as in the top panel.}
  \label{fig:Dustlayer}
\end{figure}

\subsection{RDI activation in a finite-thickness dust layer in absence of Streaming Instability}
\label{subsec:layer}

As we showed in Section~\ref{subsection:dust_density}, once the RDI is activated
by planetary waves propagating at the edges of the horseshoe region in the gas,
Streaming Instability-like perturbations arise in the dust density
(see Figs.~\ref{fig:RDIdust} and \ref{fig:rdi}). 

Nevertheless, since we considered setups in which the scale-height of dust and gas 
are initially equal ($H_\mathrm{d}=H_\mathrm{g}$) and the dust-to-gas ratio is initially
uniform ($\epsilon=0.01$) over the whole disc, other instabilities could develop due
to the rapid increase of $\epsilon$ in the disc midplane (such as Streaming Instability
or some added effect from Dust Settling Instability). 
Therefore, in order to rule out some of these other instabilities and confirm the activation
of the RDIs by planetary waves, we mow consider a configuration with a dust layer of 
finite thickness (model RDISt0$\_$01$^{H_d}$Mp1, see Table \ref{tab:simulations}),
modeled as a diffusion process with the dust scale-height given as \citep{YL2007}:
\begin{equation}
H_\mathrm{d}=\sqrt{\frac{\alpha_\mathrm{d}}{\alpha_\mathrm{d}+St}}H_\mathrm{g},
 \label{eq:Hd}
\end{equation}
where $\alpha_\mathrm{d}=10^{-4}$ is a dimensionless measurement parameter of the dust diffusion
by gas turbulence \citep[see][and references therein]{Chametla2025}. 

Equation~(\ref{eq:Hd}) assumes a turbulence model whose outcome is to facilitate the 
diffusion of the Streaming Instability in the vertical direction.
In this case, we consider that $St=0.01$, therefore we have that the dust scale-height
is $H_\mathrm{d}\approx0.1H_\mathrm{g}$. Additionally\footnote{In this model we also include
a loud white noise in the radial and vertical components of the gas and dust velocities
as in the RDISt0$\_$2Mp0 model.},
we follow \citep{BLD2021} and prescribe a vertical dust distribution such that an initially
uniform, low dust-to-gas mass ratio $\epsilon=10^{-3}$ at the midplane exponentially
decreases in the vertical direction. Overall, the setup of model RDISt0$\_$01$^{H_d}$Mp1
is designed to avoid the growth of the SI or DSI since, with a low dust-to-gas mass ratio
and a small Stokes number in a finite-thickness dust layer such instabilities do not develop
\citep[for SI criterion see][]{CL2020}.

In Fig.~\ref{fig:Dustlayer}, we show the results of this experiment. Again, we find
the same pattern in the perturbation of the dust density 
(see top panel of Fig.~\ref{fig:Dustlayer}). It should be noted that in this case
we have smoothly introduced the gravitational potential of the planet during the first
five orbits (see Appendix~\ref{sec:appendixB}), and that we use a smaller Stokes number
compared to the models discussed above. Nonetheless, in the $2D$-vorticity map
(see bottom panel of Fig.~\ref{fig:Dustlayer}) one can see that the planetary waves
start to propagate at the upstream and downstream regions. In addition, one can observe
the formation of the pw-stripes at the boundaries of the horseshoe region,
close to the planet. Note that, because the dust is initially confined to a finite layer
and is strongly coupled to the gas ($St=0.01$), it does not produce strong vertical
oscillations in the differential gas elements that make the U-turns behind (in front)
of the planet. In fact, the stripes resulting from these vertical oscillations remain
confined to the horseshoe region.

In light of the above discussion, one may expect that the development of the RDI be hindered.
Nevertheless, the dust density perturbations show the same pattern observed at early times
in the previously presented models with $M_\mathrm{p}=1M_\oplus$ and $St\neq0$.
It is important to note that the stripes confined within the horseshoe region do not produce
any disturbance in the dust density. For instance, see point $(r,\phi)=(1.0,-0.75)$ 
in Fig.~\ref{fig:Dustlayer}, which is within the upstream region on the dust density
and vortensity maps. It is clear that in this point there are disturbances in vortensity
but not in the density of the dust. Therefore, with this numerical experiment we can ruled
out that the perturbations in the dust density are the product of the SI (or DSI).
Additionally, we can argue that it is very likely that the stripes produced by
the vertical oscillations of the gas elements are not the cause of the formation
of Streaming-like perturbations either.

\section{Conclusions}
\label{sec:conclusions}

We performed 3D, high-resolution two-fluid simulations of dusty discs with embedded
low-mass planets. We found that the planetary waves (PWs), which are excited 
by the planet and propagate as vortensity disturbances in the gas, activate 
the resonant drag instability (RDI). This process is driven by a resonant interaction
with the relative streaming motion of dust and gas. As a result,
filamentary perturbation develops in the dust density, starting within the horseshoe 
region and propagating beyond it, after several orbital times. 
Since our disc is inviscid, loosely coupled dust grains undergo fast settling towards
the midplane. However, we verified that the local increase of the dust-to-gas mass ratio
alone does not lead to the RDI; the disc has to be perturbed by the planet for
the instability to develop.
The growth of the instability does not seem to strongly depend on the Stokes number
of dust grains, although the range of Stokes numbers considered in this work,
$0.01$--$0.5$, is relatively narrow.

Remarkably, we find perturbations in the vertical component of the gas velocity
of the type generated by bouyancy resonances in adibatic gaseous discs. This occurrence
suggests that, in global simulations of gas-dust discs, the dynamics of an adiabatic
gaseous disc is similar to that of an isothermal dust-gas disc with cooling in the Epstein regime
(i.e., for small Stokes numbers), as previously noted by \citet{LY2017}.

Our results suggest that low-mass protoplanets may trigger the activation of dust-gas
instabilities, thus facilitating dust clumping and planetesimal formation
in protoplanetary discs.
Such planet-induced instabilities might provide alternative pathways to the formation
of planetesimals in regions where the planetary embryos either formed or were delivered
by disc-driven migration.

\section*{Acknowledgements}

We are grateful to the referee for her/his constructive and careful
report. We thank Ondrej Chrenko and Fr\'{e}d\'{e}ric Masset for useful suggestions and valuable contributions in the first version of this manuscript. The work of R.O.C. was supported by the Czech Science Foundation (grant 21-23067M). Computational resources were available thanks to the Ministry of Education, Youth and Sports of the Czech Republic through the e-INFRA CZ (ID:90254). 
Y.H. was supported by the Jet Propulsion Laboratory, California Institute of Technology, under a contract with the National Aeronautics and Space Administration (80NM0018D0004).
G.D gratefully acknowledges support from NASA's Research Opportunities in Space and Earth Science (ROSES).

\section*{Data Availability}
The FARGO3D code is available at \href{https://github.com/FARGO3D/fargo3d}{\nolinkurl{https://github.com/FARGO3D/fargo3d}}. The input files for generating our two-fluid 3D hydrodynamical simulations will be shared on reasonable request to the corresponding author.



\bibliographystyle{mnras}
\bibliography{resonant_drag_instabilities} 




\appendix

\section{Planetary wave stripes}
\label{sec:appendixA}

In order to better identify the effect of the pw-stripes 
(hence the activation of the RDI) on the dust, we have developed 
a two-dimensional model of the generation of a planetary wave 
in the dusty-gaseous disc (without planet) described 
in Section~\ref{sec:dusty_model}, considering a Stokes number $St=0.2$.
To produce the planetary wave, we introduce the following perturbations
in the $(v_r,v_\phi)$ gas velocity components \citep[][]{Mignone2012,BLlM2016}:

\begin{equation}
   \left(
   \begin{array}{c}
               \delta v_\phi\\
               \\\delta v_r
             \end{array}\right)
             = W_A\exp\left(-\dfrac{x^2+y^2}{h_s^2}\right)\left( \begin{array}{cc}
             -\sin\phi & \cos\phi \\ 
              \cos\phi & sin\phi
             \end{array}
   \right) \left( \begin{array}{cc}
             -y \\ 
              x             \end{array}
   \right),
      \label{eq:vortex}
  \end{equation}
here, $W_A=-1$, $h_s=0.0075$ are the amplitude and size of the planetary wave, 
respectively. Note that these parameters are chosen so that the PW does not generate
additional disturbances in the gas, such as spiral waves, which could generate rapid
radial drift of the PW \citep[][]{Par2010}, as well as additional disturbances
in the dust component. 

\begin{figure*}
    \includegraphics[width=\textwidth]{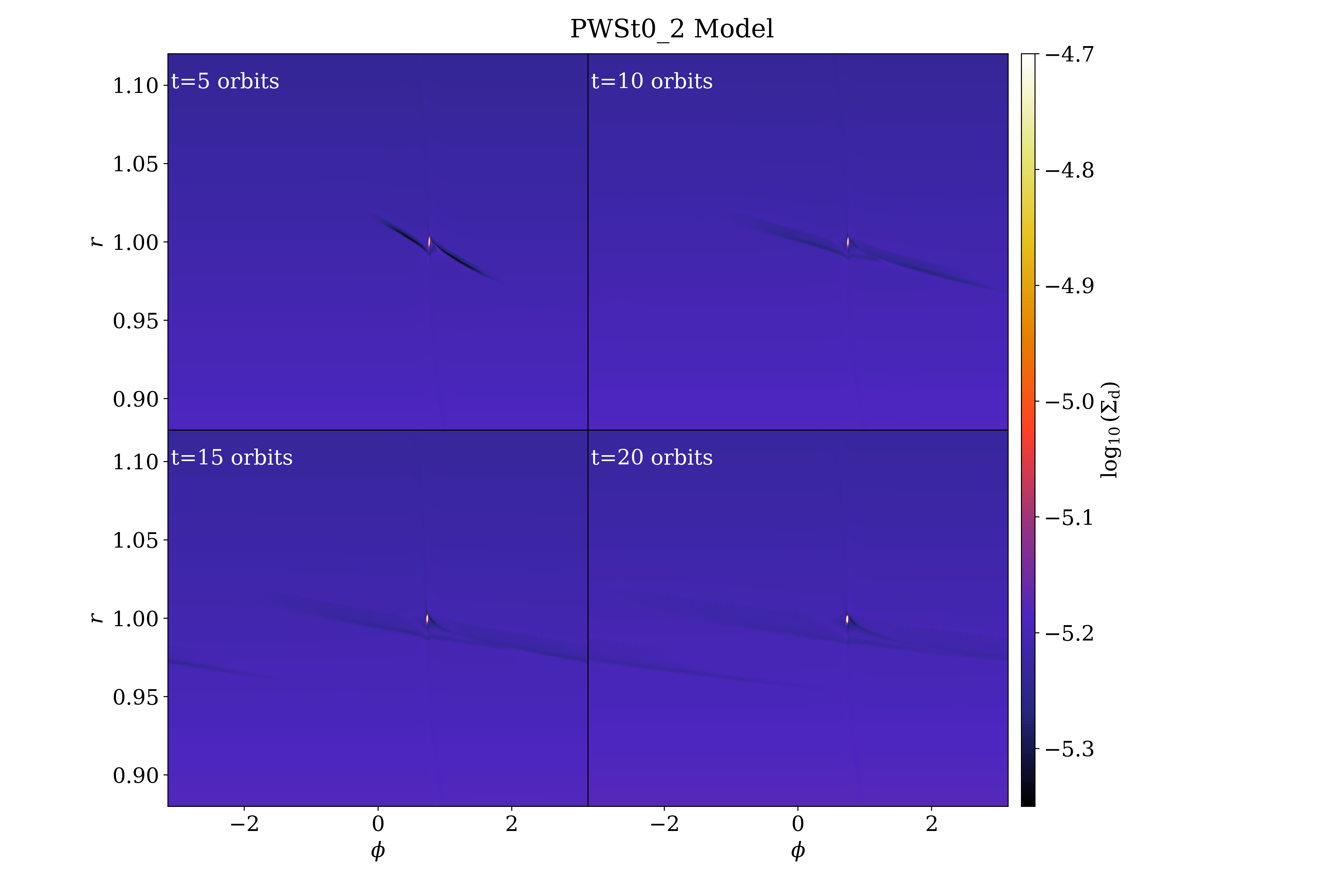}
    \caption{Temporal evolution of the planetary waves stripes in the dust density (code units) for model PWSt0$\_$2.}
    \label{fig:pws}
\end{figure*}

In Fig.~\ref{fig:pws}, we show the results of the temporal evolution 
of the dust density of model PWSt0$\_$2. One can clearly see that, at $t=5$ orbits,
the planetary wave launches two inclined stripes that start to propagate radially
and azimuthally. Note the similarity in shape and inclination with respect to
the $\phi$-axis. At later times, one can see how these opposing stripes expand
azimuthally beyond $\phi=\pm\pi/2$ until, at $t=20$ orbits, the perturbation is 
about to reach the initial position, $\phi=0$, but has expanded in the radial direction.
Something similar occurs in the density and vorticity maps discussed in the main text,
except that, in this case, many PWs produce stripes simultaneously, which inevitably
result in the filamentary perturbations reported here.

\section{Numerical implementation of the planet's gravitational potential}
\label{sec:appendixB}

\begin{figure}
  \begin{subfigure}[b]{0.49\textwidth}
  \centering
    \includegraphics[width=\textwidth]{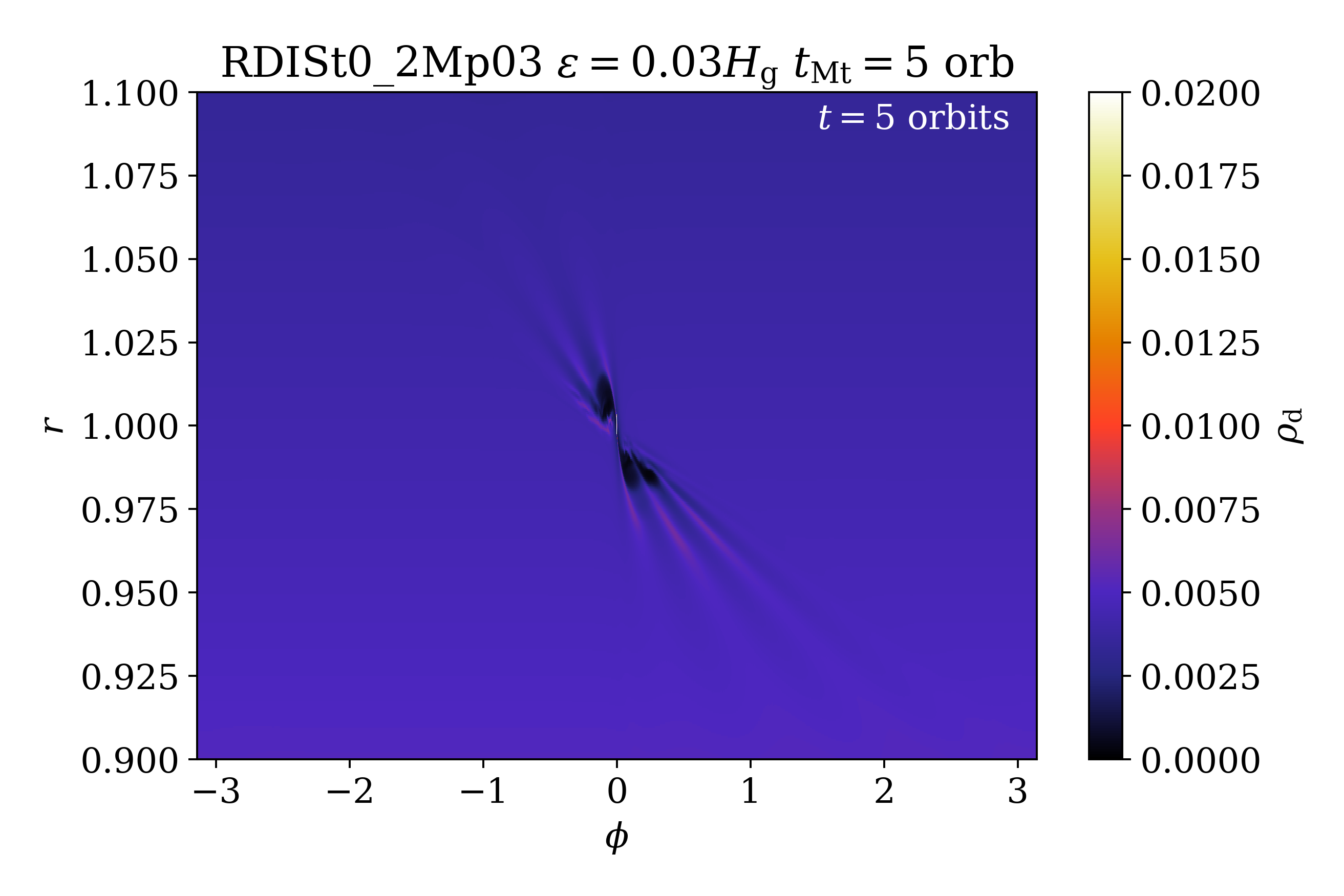}
     
  \end{subfigure}
  \hfill
  \begin{subfigure}[b]{0.49\textwidth}
  \centering
    \includegraphics[width=\textwidth]{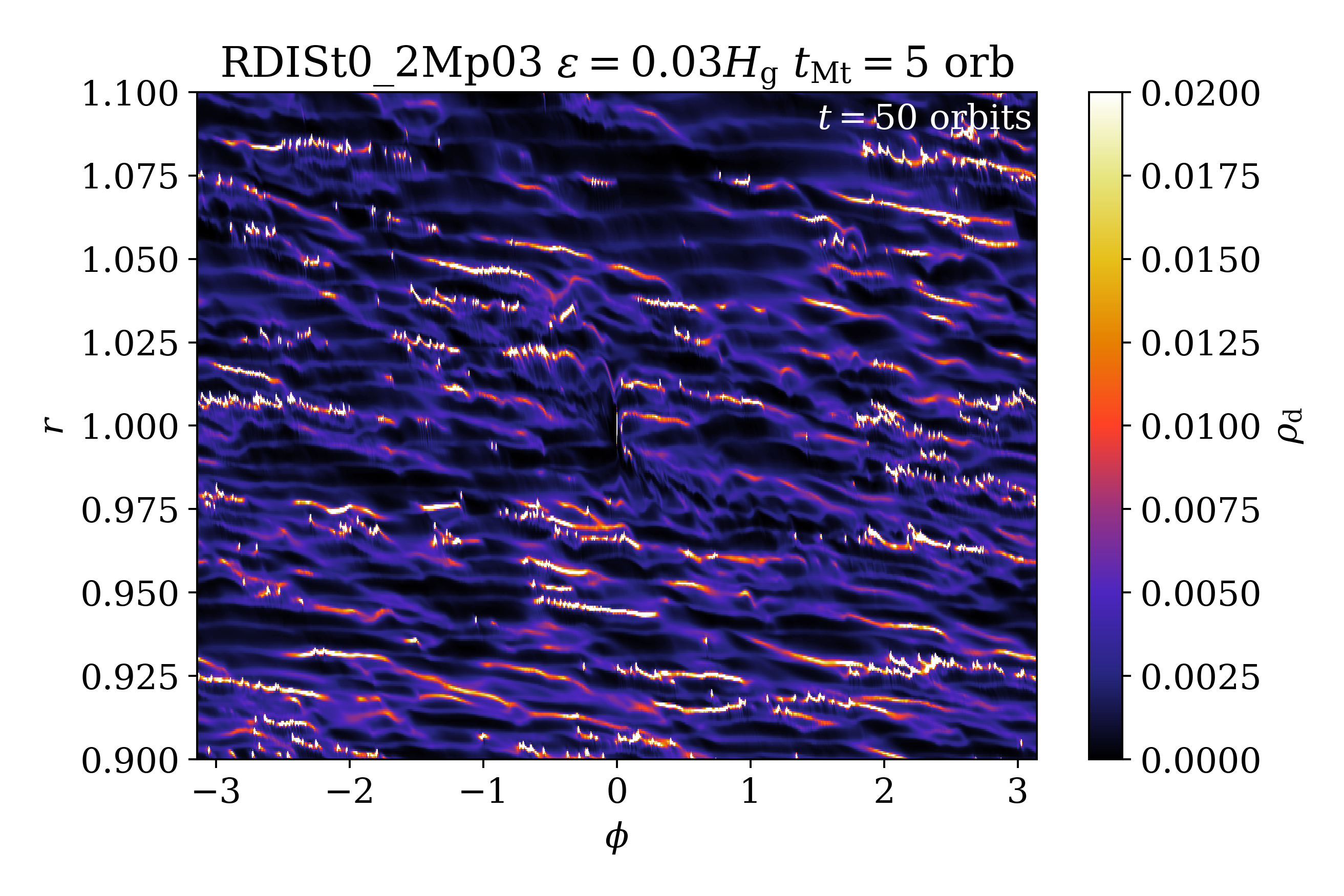}
   
  \end{subfigure}

  \caption{\textit{Top}. Dust density distribution at the midplane for the RDISt0$\_$2Mp03 model calculated at $t=5$ orbits. This model includes a larger softening length $\varepsilon=0.03H_\mathrm{g}$ and a timescale $t_\mathrm{Mt}=5$ orbits. \textit{Bottom.} Dust density distribution calculated at $t=50$ orbits for the same model shown in the top panel.}
  \label{fig:gpp}
\end{figure}

Fig.~\ref{fig:gpp} shows the dust density in the midplane of the disc for 
model RDISt0$\_$2Mp03, at $t=5$ and $t=50$ orbits, for a case in which the smoothing length
in the gravitational potential of the planet is $\varepsilon=0.03 H_\mathrm{g}$.
In this experiment, we additionally introduced the planet into the disc smoothly.
The increase of the gravitational potential of the planet was realized by means
of the mass taper function given as
\begin{equation}
  M_\mathrm{p}(t)= \left\{ \begin{array}{lcc}
             \frac{1}{2}\left[1-\cos{\left(\dfrac{\pi t}{t_\mathrm{Mt}}\right)}\right] &   \mathrm{if}  & t < t_\mathrm{Mt} \\
             \\ 1 &  & \mathrm{otherwise}, 
             \end{array}
   \right.
      \label{eq:Mp_t}
  \end{equation}
where $t_\mathrm{Mt}$ is the timescale at which the planet reaches its full mass.
Here, we consider the time $t_\mathrm{Mt}=5$ orbits. We did not find any significant changes
in the development of the resonant drag instability in the dust density when comparing with
the model in which $\varepsilon=0.01H_\mathrm{g}$ and $t_\mathrm{Mt}=0$ (see upper panel
in Fig.~\ref{fig:RDIdust}).
Therefore, contrary to 2D models \citep[see][]{OKS2015}, the introduction and the softening
of the gravitational potential of a planet does not produce artificial perturbations
in the gas dynamics around the planet. Therefore, the development of the RDI in the dust
can be considered a physical effect of the drag forces between dust and gas.

\section{Dust density distribution without feedback}
\label{sec:appendixC}

\begin{figure}
    \includegraphics[width=\columnwidth]{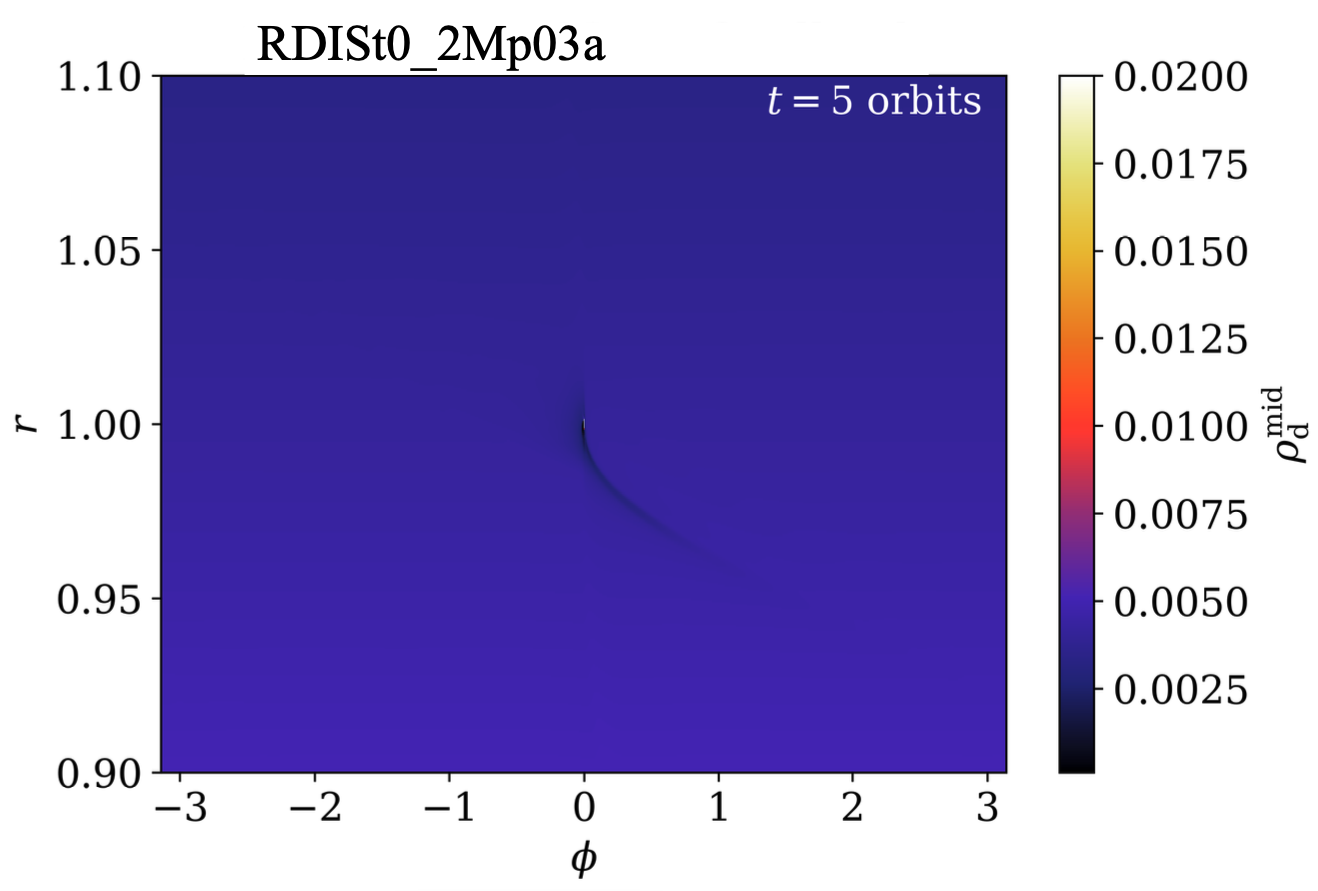}
    \caption{Dust density at the midplane for model RDISt0$\_$2Mp03a, which does not
    include dust feedback, at $t=5$ orbits.}
    \label{fig:nofeed_}
\end{figure}

In the resonant drag instability (RDI) theory approach
\citep[][]{Squire_Hopkins2018,SH2018,SH2020Physicalmodels}, the back-reaction force
of dust on gas is a key component in the formation and growth rate of
streaming instability \citep[][]{YG2005}, as well as in the formation and evolution
of structures in the protoplanetary disc, such as vortices and rings
\citep[][]{Fu2014,Dong2017, Huang2020}.
In our study, in addition to the high resolution used in our 3D numerical models,
the inclusion of such a force (frequently called feedback) is a novelty, considering
that we study the gravitational interaction of a planet using global models
of a dusty gaseous disc. To show the effects of feedback (last term on the right hand side
in Eqs.~8 and 9), in Fig.~\ref{fig:nofeed_} we show the perturbed density of the dust
in the midplane of the disc at $t=5$ orbits, for the RDISt0$\_$2Mp03a model.
In this case the feedback of the dust on the gas is not included.
This figure should be compared with Fig.~\ref{fig:gpp} (and with Fig.~\ref{fig:RDIdust}).
One can see that the dust distribution around the planet's position is completely different.
For the case without feedback, we find that the disturbance generated in the dust behave
similarly to what was previously reported in 2D numerical models
\citep[][]{BLlP2018,Regaly2020,Chrenko2024}, which did not include the effect of dust feedback.
We monitored the time evolution of the dust density up to the end of the simulations and
we did not find any sign of a streaming instability developing.

\bsp	
\label{lastpage}
\end{document}